\documentclass[prb,amsmath,amsfonts,showpacs]{revtex4}
\usepackage{graphicx}

\pdfoutput=1

\begin{document}

\title{Dynamical correlations in electronic transport through a system of coupled quantum dots}

\author{Grzegorz Micha{\l}ek}
\email{grzechal@ifmpan.poznan.pl}

\author{Bogdan R. Bu{\l}ka}

\affiliation{Institute of Molecular Physics, Polish Academy of Sciences, ul.~M.~Smoluchowskiego~17, 60-179~Pozna\'{n}, Poland}

\date{\today}

\begin{abstract}
Current auto- and cross-correlations are studied in a system of two capacitively coupled quantum dots. We are interested in a role of Coulomb interaction in dynamical correlations, which occur outside the Coulomb blockade region (for high bias). After decomposition of the current correlation functions into contributions between individual tunneling events, we can show which of them are relevant and lead to sub-/supper-Poissonian shot noise and negative/positive cross-correlations. The results are differentiated for a weak and strong inter-dot coupling. Interesting results are for the strong coupling case when electron transfer in one of the channel is strongly correlated with charge drag in the second channel. We show that cross-correlations are non-monotonic functions of bias voltage and they are in general negative (except some cases with asymmetric tunnel resistances). This is effect of local potential fluctuations correlated by Coulomb interaction, which mimics the Pauli exclusion principle.
\end{abstract}

\maketitle

\section{Introduction}

Recently, McClure et al.\cite{mcclure07} and Zhang et al.\cite{zhang07} performed measurements of current noise auto- and cross-correlation in a system of two capacitively coupled quantum dots (2QD). It is first measurement showing some aspects of current correlations in 2QD, as anti-bunching and bunching of scattered electrons. The 2QD system is simplest, in which one can see a role of Coulombic interactions on current-current correlations.

In a system of noninteracting electrons the Pauli principle is essential in scattering process and it can lead to anti-bunching.\cite{feynman} This effect in a multiterminal geometry, such as in the Hanbury Brown and Twiss experiment,\cite{hbt} can be manifested in negative cross-correlation between scattered electrons. The phenomena was studied theoretically\cite{loudon,texier,martin} and observed in several experiments.\cite{liu,oliver,henny} In nanostructures anti-bunching is responsible for reduction of auto-correlation shot noise $S_{II}$ below the Poissonian value $S_P=2eI$, where $I$ is the average current and $e$ is the charge of an electron.\cite{kulik,khlus,lesovik,buttiker1990,blanter} For sequential tunneling through a single quantum dot (QD) a maximal reduction of the shot noise can be to the value $1/2\times 2eI$.\cite{korotkov92,hershfield,korotkov94,blanter}

An enhancement of shot noise above the Poissonian value was observed in a resonant tunneling diode by Iannaccone et al.\cite{iannaccone} and in a quantum well by Kuznetsov et al.\cite{kuznetsov} in the region of negative differential resistance. The enhancement of shot noise was interpreted as a result of change in the density of states, which causes more state to be available for tunneling from cathode. Strong back-scattering, at current pinch-off (close to the Coulomb blockade border), can be also responsible for an enhancement of shot noise.\cite{bb99,safonov} In the Coulomb blockade region inelastic cotunneling processes can affect transport and lead to super-Poissonian shot noise.\cite{sukhorukov,thielmann}

Recently, Wu and Yip,\cite{wu} and Rychkov and B\"{u}ttiker\cite{rychkov} applied a phenomenological Langevin formalism in circuit modeling to calculations of current cross-correlations in a 3-terminal system. An interesting result is that current cross-correlations are always positive in a macroscopic classical 3-terminal with a fluctuating current in an input electrode. It is in contrast to a microscopic picture of scattered electrons, which in general shows anti-bunching with negative cross-correlations.\cite{loudon,texier,martin} However, there is an exception when inelastic scattering occurs. Texier and B\"{u}ttiker\cite{texier} considered a multiterminal system with inelastic scattering in an additional electrode, in which the current was kept equal to zero and which caused voltage fluctuations. They showed that inelastic scattering can lead to positive cross-correlations, whereas for quasielastic scattering correlations remain always negative. Just recently Oberholzer at al.\cite{oberholzer} confirmed experimentally these predictions in a device with fractional quantum Hall effect edge states, in which interactions between current carrying states and a fluctuating voltage were controlled by an external gate voltage. If scattered electrons are entangled, the situation can be different, one can see bunching or anti-bunching. Burkard et al.\cite{burkard} showed that for singlets the auto-correlation function is twice as large as for independent electrons and is reduced to zero for triplets.

Current correlations in the system of two capacitively coupled quantum dots were studied theoretically in several papers.\cite{michalek,aghassi08,haupt08,hwang07,weymann08} Our studies on a ferromagnetic single electron transistor (fSET)\cite{bb99,bb00} are closely related, because there are two channels for electrons with opposite spins $\sigma=\uparrow$, $\downarrow$ and Coulomb interactions are included as well. For the first time we pointed on dynamical Coulomb blockade effect between the both channels, which is responsible for a reduction of current and leads to the super-Poissonian shot noise. Positive cross-correlations were predicted by Cottet et al.\cite{cottetprb,cottetprl,cottetepl} on a similar system with three terminals and 3QDs. They also analyzed time evolution of tunneling events and presented bunching. Dynamical Coulomb blockade and bunching were studied by Gustavsson et al.\cite{gustavssonprl,gustavssonprb} in time-resolved measurements of electron transport through a multilevel QD.

In this paper we want to study a microscopic nature of dynamical current correlations, especially a role of Coulomb interactions, in the 2QD system. Some problems were mentioned earlier,\cite{michalek,bb00} for example, a role of charge and polarization fluctuations and their contribution to the current correlation function. We expect that for a symmetric case these contributions may compensate each other, but they can be pronounced in an asymmetric case. Two different situations will be studied, with a weak and strong inter-dot coupling. We will show that both the cases have different charging diagrams in the space of biases applied to the both channels. For the strong coupling electron transfer in one of the channel should be strongly correlated with charge drag in the second channel. In order to see correlations between individual tunneling events we decompose correlation functions and analyze contributions leading to the sub-/super-Poissonian noise as well to positive cross-correlations. The experiments\cite{mcclure07,zhang07} were performed for small biases and near degeneracy lines separating different Coulomb blockade regions. The present studies are more general. We analyze the Schottky contribution, which can be large close to the Coulomb blockade border due to backward scattering. However, our main interest focus on dynamical correlations for electronic transport outside the Coulomb blockade region when the bias window is large.

The paper is organized as follows. In the next section II we describe the 2QD model, derive general formulas for charging energy, current and correlation functions. The section III presents charging diagrams and allowed tunneling processes in the space of bias voltages. The charging diagrams are background for further analysis of correlations between individual tunneling events presented in the section IV. We show there differences of auto- and cross-correlations for weak and strong inter-dot coupling. The correlation functions are non-monotonous with bias voltages, which means that there are several competing correlation processes. Our decomposition approach shows which contributions are relevant. For example, current cross-correlations are in general negative (except special situations with asymmetric resistances), but this anti-bunching effect is due to potential fluctuations (it is not due to the Pauli exclusion principle as in many nanostructures\cite{kulik,khlus,lesovik,buttiker1990,blanter}). Main results of the paper are summarized in the final section V.

\section{Description of the model and general derivation of transport quantities}

\subsection{Model}

\begin{figure}
\includegraphics[width=0.4\textwidth]{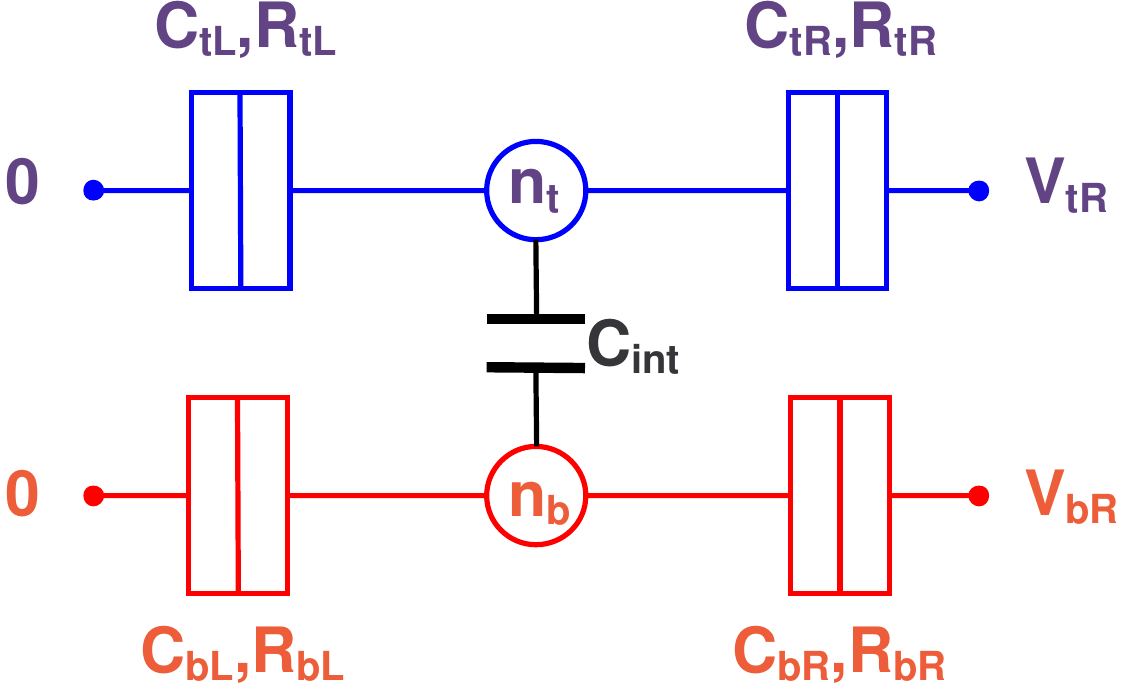}
\caption{(Color online) Schematic view of two capacitively coupled quantum dots. Each dot $\ell$ ($\ell={t,b}$ -- top, bottom) is connected with the electrodes ($\alpha=L,\,R$ -- left, right). The tunnel junctions are characterized by the capacitances $C_{\ell\alpha}$ and the resistances $R_{\ell\alpha}$. The strength of the dot-dot interaction is modeled by the capacitance $C_{int}$. The bias voltages $V_{tR}, V_{bR}$ are applied to the right electrodes.}\label{fig1}
\end{figure}

We consider a system which consists of two large capacitively coupled quantum dots (top and bottom: $\ell=t,\,b$) with a continuous electronic density of states. Each quantum dot (QD) is connected with the left ($\alpha=L$) and the right ($\alpha=R$) electrodes through the tunnel junctions which are
characterized by the tunnel resistances $R_{\ell\alpha}$ and capacitances $C_{\ell\alpha}$ (see Fig.~\ref{fig1}). The strength of the inter-dot Coulomb interaction is modeled by the capacitance $C_{int}$. To this 4-terminal system, the bias voltages $V_{tR}$, $V_{bR}$ are applied to the right electrodes, while both the left electrodes are grounded (asymmetric bias). Usually additional gates are applied to the QDs, which can change their electrochemical potentials and number of electrons. In order to simplify the analysis these gates can be omitted. The corresponding resistances of the tunnel junctions are assumed to be much larger than quantum resistance $R_\mathrm{Q} = h/2e^2$  so the electronic transport is dominated by incoherent, sequential tunneling processes.\cite{schon} Since our research is focused on dynamical processes in high voltage regime (outside the Coulomb blockade), higher order processes (e.g., cotunneling) are neglected. Cotunneling can be relevant in the Coulomb blockade region, for tunneling currents as well as super-Poissonian shot noise.\cite{aghassi08,haupt08,hwang07,weymann08} It is also assumed that the temperature $T$ is low which means that charging energies of the individual dots $E^{ch}_{\ell}$ and an energy of the inter-dot Coulomb interaction $E^{ch}_{int}$ are much larger than a thermal energy $k_B T$, i.e., $E^{ch}_{\ell}, E^{ch}_{int} \gg k_B T$.

\subsection{Charging energy}

The voltage drops on the tunnel junctions as well as between both dots (i.e., on the capacitance $C_{int}$) can be calculated from Kirchhoff's laws:
\begin{equation}
V^{drop}_{tL}= \frac{E^{ch}_t}{e}n_t + \frac{E^{ch}_{int}}{e}n_b + \frac{C_{tR} C_b}{C^2}V_{tR} + \frac{C_{bR} C_{int}}{C^2}V_{bR}
\end{equation}
\begin{equation}
V^{drop}_{tR} = V_{tR} - V^{drop}_{tL}
\end{equation}
\begin{equation}
V^{drop}_{bL} = \frac{E^{ch}_{int}}{e}n_t + \frac{E^{ch}_b}{e}n_b + \frac{C_{tR} C_{int}}{C^2}V_{tR} + \frac{C_t C_{bR}}{C^2}V_{bR}
\end{equation}
\begin{equation}
V^{drop}_{bR} = V_{bR} - V^{drop}_{bL}
\end{equation}
\begin{equation}
V^{drop}_{int} = \frac{e(C_{bL} + C_{bR})}{C^2}n_t - \frac{e(C_{tL} + C_{tR})}{C^2}n_b + \frac{(C_{bL} + C_{bR})C_{tR}}{C^2}V_{tR} - \frac{(C_{tL} +
C_{tR})C_{bR}}{C^2}V_{bR}
\end{equation}
where an electron charge $e<0$, $C_t=C_{tL}+C_{tR}+C_{int}$, $C_b=C_{bL}+C_{bR}+C_{int}$, $C^2=C_t C_b-C^2_{int}$. The charging energies
$E^{ch}_t=e^2 C_b/C^2$, $E^{ch}_b=e^2 C_t/C^2$ and $E^{ch}_{int}=e^2 C_{int}/C^2$. The inter-dot energy $E^{ch}_{int}$ describes the change in the energy of one dot when an electron is added to the other dot.

Transfer rate $\Gamma^s_{\ell\alpha}$ through the junction $\ell\alpha$ can be found from a changes of the free energy of the whole system when an electron is added ($s=+$) (or extracted: $s=-$) to (from) one of the quantum dots. The free energy consists of the potential energies of the electrodes and the electrostatic energies of the charged capacitors and for the system under consideration has a form:
\begin{eqnarray}
F(n_t,n_b,n_{tL},n_{tR},n_{bL},n_{bR})=- Q_{tRel}V_{tR} - Q_{bRel}V_{bR} \nonumber\\
+\left[C_{tL}(V^{drop}_{tL})^2 + C_{tR}(V^{drop}_{tR})^2 + C_{bL}(V^{drop}_{bL})^2 + C_{bR}(V^{drop}_{bR})^2 + C_{int}(V^{drop}_{int})^2\right]/2\;,
\end{eqnarray}
where $Q_{tRel} = Q_{tR} + en_{tR}$ and $Q_{bRel} = Q_{bR} + en_{bR}$ are charges on the top and bottom right electrodes, respectively. $n_{tR}$ ($n_{bR}$) is a number of electrons transferred from the right electrode to the top (bottom) QD and charges on the capacitors $C_{\ell\alpha}$ are
$Q_{\ell\alpha} = C_{\ell\alpha}V^{drop}_{\ell\alpha}$ ($\ell=t,\,b$; $\alpha=L,\,R$). The changes of the free energy due to electron tunneling
through the junction $\ell\alpha$ can be written as a differences between the free energies of the initial $F_i$ and the final $F_f$ states
\begin{eqnarray}
\Delta F^s_{tL}(n_t,n_b) = F_i(n_t, n_b, n_{tL}, n_{bL}, n_{tR}, n_{bR}) - F_f(n_t +s 1, n_b, n_{tL}+s1, n_{bL}, n_{tR}, n_{bR})\,,\\
 \Delta
F^s_{bL}(n_t,n_b) = F_i(n_t, n_b, n_{tL}, n_{bL}, n_{tR}, n_{bR}) - F_f(n_t, n_b+ s1, n_{tL}, n_{bL} +s1, n_{tR}, n_{bR})\,,\\
 \Delta
F^s_{tR}(n_t,n_b) = F_i(n_t, n_b, n_{tL}, n_{bL}, n_{tR}, n_{bR}) - F_f(n_t +s1, n_b, n_{tL}, n_{bL}, n_{tR}+ s1, n_{bR})\,,\\
 \Delta
F^s_{bR}(n_t,n_b) = F_i(n_t, n_b, n_{tL}, n_{bL}, n_{tR}, n_{bR}) - F_f(n_t, n_b+ s1, n_{tL}, n_{bL}, n_{tR}, n_{bR}+ s1)\,,
\end{eqnarray}
where $s=\pm$ for an added or extracted electron. An electron can be transferred through the junction $\ell\alpha$ when the corresponding free energy difference $\Delta F^s_{\ell\alpha}(n_t,n_b) > 0$.

\subsection{Current and correlation functions}

One can calculate the tunneling rates through the junctions using the method developed for a single QD with a continuous electronic density of states (DOS) (see, for example,\cite{schon}). For our system, one can obtain the following tunneling rate for transfer of an electron through the junction $\ell\alpha$:
\begin{equation}\label{gamma}
\Gamma^s_{\ell\alpha} (n_t,n_b) = \frac{1}{e^2 R_{\ell\alpha}} \frac{\Delta F^s_{\ell\alpha}(n_t,n_b)}{1 - \exp\left[-\Delta F^s_{\ell\alpha}(n_t,n_b)/k_BT\right]}\,,
\end{equation}
where the tunneling resistance $R_{\ell\alpha}= \hbar/(4 \pi e^2 |M_{\ell\alpha}|^2
D_\ell D_\alpha)$ and $D_\ell$ is DOS in QD, while $D_\alpha$ is DOS in the electrode. It is assumed that the transfer matrix element $M_{\ell\alpha}$ as well as DOS are constant around the Fermi energy, and the resistances  $R_{\ell\alpha}$ are parameters of the model.

In the stationary state the average currents flowing through the junction $\ell\alpha$ can be found from the formula
\begin{equation}\label{current}
I_{\ell\alpha} = (\delta_{L\alpha}-\delta_{R\alpha})(I^+_{\ell\alpha}-I^-_{\ell\alpha}) \,,
\end{equation}
where  $\delta_{L\alpha}$, $\delta_{R\alpha}$ are the Kronecker's deltas, $I^s_{\ell\alpha}=e\sum_{n_t,n_b}\Gamma^s_{\ell\alpha}(n_t,n_b)\; p(n_t,n_b)$ is the current flowing to/from the QD ($s=\pm$), and the probability $p(n_t,n_b)$ describes the system in the steady state which contains $n_t$ and $n_b$ excess electrons on the top and the bottom QD, respectively. The probability $p(n_t,n_b)$ can be found from the master equation:
\begin{eqnarray}\label{master}
\frac{d p(n_t,n_b;t)}{dt}=\sum_{\alpha,s}\left[\Gamma^s_{t\alpha}(n_t-s1,n_b)p(n_t-s1,n_b;t)+ \Gamma^s_{b\alpha}(n_t,n_b-s1)p(n_t,n_b-s1;t)\right]\nonumber\\
-p(n_t,n_b;t)\sum_{\alpha,s}\left[\Gamma^s_{t\alpha}(n_t,n_b)+\Gamma^s_{b\alpha}(n_t,n_b) \right]
\end{eqnarray}
with the left hand side equal to zero.

To analyze fluctuations in the system we extend the generation-recombination approach\cite{vliet} for multi-electron channels by a generalization of the method developed for spinless electrons in a SET.\cite{korotkov94} In calculation of the current-current correlation functions, we also include the self-correlation terms as well.\cite{korotkov94} According to this procedure the auto- and cross- current-current correlation functions are
\begin{equation}\label{noise}
S_{\ell\alpha,\ell'\alpha'}(\omega) = \delta_{\ell\ell'}\delta_{\alpha\alpha'}S^{Sch}_{\ell\alpha} + S^c_{\ell\alpha,\ell'\alpha'}(\omega),
\end{equation}
where the Schottky term (the frequency independent part for $\omega \rightarrow \infty$) is given by
\begin{equation}\label{sch}
S^{Sch}_{\ell\alpha} = 2e(I^+_{\ell\alpha}+I^-_{\ell\alpha})\,,
\end{equation}
and the frequency dependent part
\begin{eqnarray}\label{sc}
S^c_{\ell\alpha,\ell'\alpha'}(\omega)= 2e^2\,(\delta_{L\alpha}-\delta_{R\alpha})(\delta_{L\alpha'}-\delta_{R\alpha'})\sum_{n'_t,n'_b;n_t,n_b} D_{\ell\alpha,\ell'\alpha'}^{n'_tn'_b,n_tn_b}(\omega)\,.
\end{eqnarray}
Here, we denoted
\begin{eqnarray}\label{scdyn}
D_{\ell\alpha,\ell'\alpha'}^{n'_tn'_b,n_tn_b}(\omega)= \sum_{s,s'}ss'\,\Bigl\{\Gamma^{s'}_{\ell'\alpha'}(n'_t,n'_b)\,G_{n'_tn'_b,n_tn_b}(\omega)
\bigl[\delta_{t\ell}\Gamma^s_{\ell\alpha}(n_t-s1,n_b)\,p(n_t-s1,n_b)\nonumber\\
+\delta_{b\ell}\Gamma^s_{\ell\alpha}(n_t,n_b-s1)\,p(n_t,n_b-s1)\,\bigr]\nonumber\\
+\Gamma^{s}_{\ell\alpha}(n'_t,n'_b)\,G_{n'_tn'_b,n_tn_b}(-\omega) \bigl[\delta_{t\ell'}\Gamma^{s'}_{\ell'\alpha'}(n_t-s'1,n_b)\,p(n_t -s'1,n_b)\nonumber\\+\delta_{b\ell'}\Gamma^{s'}_{\ell'\alpha'}(n_t,n_b-s'1)\,p(n_t,n_b-s'1)\,\bigr]\Bigr\}
\end{eqnarray}
as contributions to the dynamical part of the correlation function for various tunneling events in the space of the charge states $(n_t,n_b)$. Later we will analyze the components $D_{\ell\alpha,\ell'\alpha'}^{n'_tn'_b,n_tn_b}$, in order to show correlations between tunneling processes. The elements of the Green function $G_{n'_tn'_b,n_tn_b}(\omega)=(i\omega\hat{1} -\hat{M})^{-1}_{n'_tn'_b,n_tn_b}$ and the matrix $\hat{M}$ is constructed from the right hand side of the master equation (\ref{master}). Here, we use the symmetrized formulas [Eqs.~(\ref{sc})-(\ref{scdyn})] for the current shot noise. Recently, an asymmetric formula was used to studies frequency-dependent asymmetric features.\cite{symmetric} Our analysis is confined to the zero-frequency limit, and therefore, the symmetrized approach is justified.

We can write also other correlation function for any quantity $X$ and $Y$ in the space of states $(n_t,n_b)$
\begin{eqnarray}\label{sxy}
S_{XY}(\omega) = 4 \sum_{n'_t,n'_b;n_t,n_b} X(n'_t,n'_b)
G_{n'_tn'_b,n_tn_b}(\omega) Y(n_t,n_b) p(n_t,n_b)\,. \end{eqnarray}
For example, in further studies we will calculate the voltage-voltage correlation function $S_{V_{t},V_{t}}$, where $V_{t}=(E^{ch}_t n_{t}+E^{ch}_{int} n_b)/e$ and it is a potential difference at the top QD when an additional charge is added to the system. The function $S_{V_{t},V_{t}}$ describes potential and charge fluctuations during current flow.

\section{Charging diagrams}

Before solving the master equation (\ref{master}), we should first to determine [from Eq.~(\ref{gamma})]  the tunneling rates $\Gamma^s_{\ell\alpha} (n_t,n_b)$ in the two-dimensional (2D) space of charge states $\{n_t,n_b\}$ for given bias voltages $V_{tR}$ and $V_{bR}$. If $\Delta F^s_{\ell\alpha}(n_t,n_b)<0$ then $\Gamma^s_{\ell\alpha} (n_t,n_b)\approx \exp\left[-|\Delta F^s_{\ell\alpha}(n_t,n_b)|/k_BT\right]/(e^2R_{\ell\alpha})$ and the tunneling process is exponentially suppressed in the low temperature limit ($T\to 0$). Otherwise, when $\Delta F^s_{\ell\alpha}(n_t,n_b)> 0$, the tunneling process can be relevant for transport. In practice one can restrict considerations to the $\{n_t,n_b\}$ space around the ground state $(0,0)$ with a small number of available states and tunneling processes relevant for transport. Of course, the space of available states depends on the bias voltage $V_{\ell\alpha}$ and increases with an increase of $V_{\ell\alpha}$. If one wants to consider temperature effects, the space $\{n_t,n_b\}$ should be enlarged and neighboring charge states with thermally activated processes should be taken into account.

\begin{figure}
\includegraphics[width=0.4\textwidth]{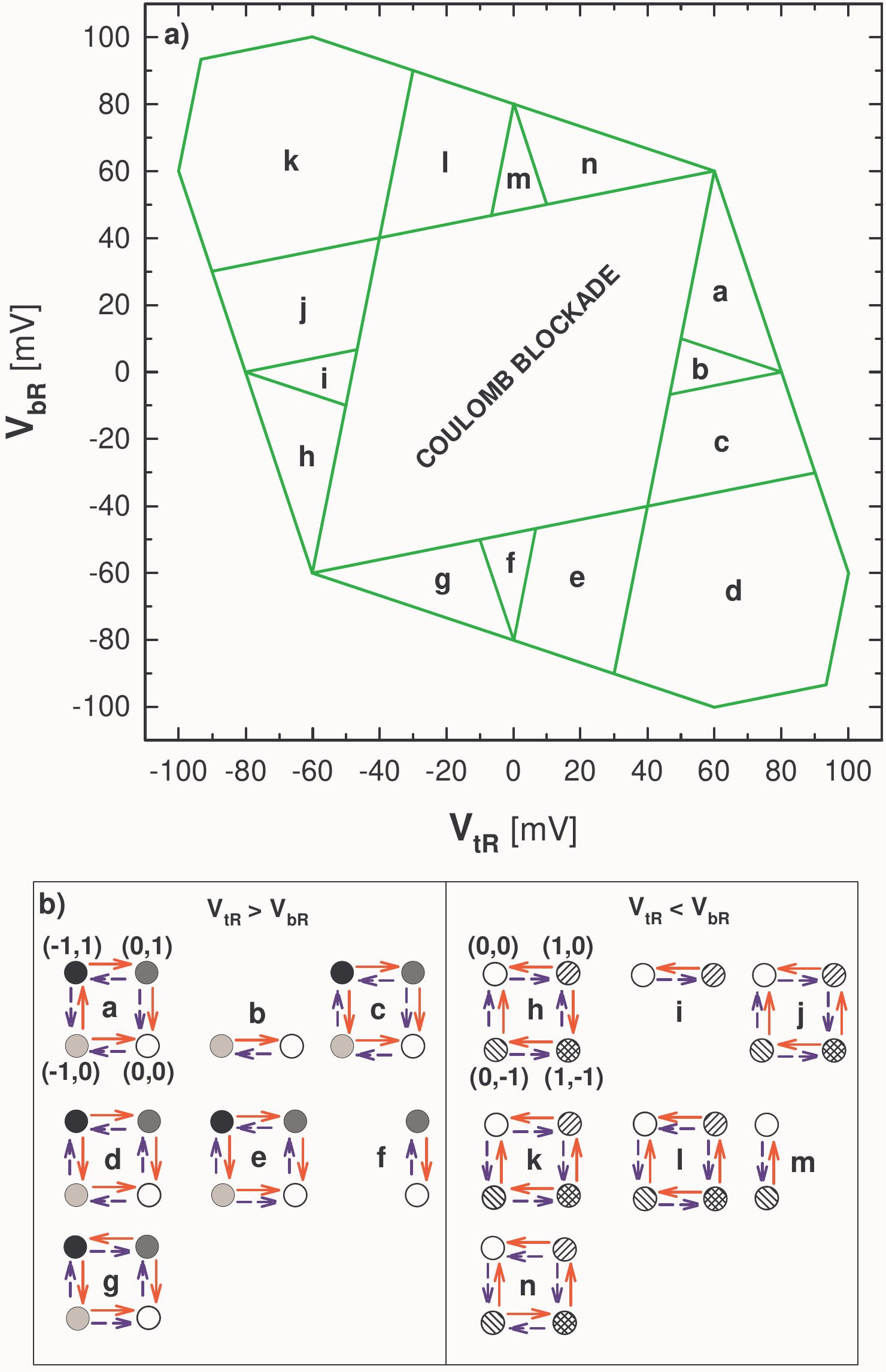}
 \caption{(Color online) (a) The diagram of the available charge states and tunneling processes contributing to transport in the space of bias voltages $(V_{tR}, V_{bR})$ for the coupling $C_{int} = C_{\ell\alpha}= 1$ aF. Solid (green) lines denote threshold voltages $V^{th}_{\ell\alpha}$, which set regions $\{ a,\ldots,n \}$ with particular allowed tunneling processes $\{ \Gamma^s_{\ell\alpha}(n_t,n_b) \}$. b) Diagrammatic representation of the tunneling processes $\{ \Gamma^s_{\ell\alpha}(n_t,n_b) \}$ between the charge states $(n_t,n_b)$ allowed in transport for various regions $\{ a,\ldots,n \}$. Tunneling processes between charge states are shown as (red) solid arrows for tunneling through the left junction and (blue) dash arrows denote tunneling through the right junction; horizontal arrows correspond to tunneling processes through the top QD, whereas vertical arrows -- for tunneling through the bottom QD, respectively. For example, the diagram ``$b$'' presents allowed tunneling processes between states $(0,0)$ and $(-1,0)$ through the left and the right junction in the top QD. The solid (red) arrow between states $(-1,0)$ and $(0,0)$ denotes $\Gamma^+_{tL}(-1,0)$, while the dash (blue) line  between these states corresponds to $\Gamma^-_{tR}(0,0)$. The other processes in the regions $\{b,...,n\}$ can be described analogously. For larger voltages $|V_{tR}|$ and $|V_{bR}|$ (i.e., outside the regions enclosed by the green lines) additional new charge states appears, e.g., $(2,0)$, $(-2,1)$, etc.}\label{fig2}
\end{figure}

Here, we perform an analysis of the available states and relevant tunneling processes at $T=0$ for a weak and strong inter-dot coupling case. First, the weak coupling case is considered, for which one can expect that transport is similar to the case of two independent QDs and one can easily understand modifications introduced by the inter-dot coupling. We consider the system with symmetrical couplings to the electrodes $C_{tL}=C_{tR}=C_{bL}=C_{bR}$, in which the charging energy of the both QDs is  $E^{ch}_t=E^{ch}_b$. This case is simpler to presentation. It is no problem to generalize the analysis for different couplings $C_{\ell\alpha}$. The available charge states $\{n_t,n_b\}$ and tunneling processes participating in currents are presented in Fig.~\ref{fig2} in the ($V_{tR}$,$V_{bR})$ space. The diagram was obtained for $C_{int} = C_{\ell\alpha} = 1$ aF. The rhombic area in the middle of the Fig.~\ref{fig2}(a) corresponds to the Coulomb blockade (CB) region, where all $\Gamma^-_{\ell\alpha}(0,0)$ are suppressed and electrons can not leave the state $(0,0)$. With increasing voltage $V_{tR}$ (or $V_{bR}$) some tunneling processes become allowed, because the corresponding free energy differences become positive, $\Delta F^s_{\ell\alpha}(0,0)>0$. New charge states $(n_t,n_b)$ become available and the currents begin to flow. Threshold voltages $V^{th}_{\ell\alpha}$ can be determined from the equation $\Delta F^s_{\ell\alpha}(n_t,n_b)=0$ and they are presented in Fig.~\ref{fig2}(a) as (green) solid lines. The threshold lines divide the $(V_{tR},V_{bR})$ space into regions $\{ a,\ldots,n \}$ with particular allowed tunneling processes $\{ \Gamma^s_{\ell\alpha}(n_t,n_b) \}$ presented below in Fig.~\ref{fig2}(b). Let us analyze as an example transport for the case of small $V_{bR}\approx 0$. With an increase of $V_{tR}$ one can reach above a threshold voltage $V^{th}$ the region $b$, where $\Gamma^-_{tR}(0,0)$ becomes relevant and electron can leave the top QD through the right tunnel junction. In this case only two charge states $(0,0)$ and $(-1,0)$ participate in transport. In this region one can expect similar current and shot noise characteristics as those for independent QD. The tunneling current is flowing only through the top QD while the bottom QD is still in the Coulomb blockade regime. On the other hand, in the region $f$ only two charge states $(0,0)$ and $(0,1)$ participate in transport so the tunneling current is flowing through the bottom QD while the top QD is in the CB regime. In the region $a$ the voltage $V_{bR}$ is higher, $\Gamma^-_{bL}(-1,0)$ becomes relevant (tunneling through the bottom left junction is allowed) and four charge states $(0,0)$, $(-1,0)$, $(-1,1)$, $(0,1)$ participate in transport. On the left hand side of Fig.~\ref{fig2}(a) we have $V_{tR}<V_{bR}$ and the situation is symmetric to that one on the right hand side. For higher voltages, outside the regions $\{ a,\ldots,n \}$, additional states $(n_t,n_b)$ are available for the tunneling events.

\begin{figure}
\includegraphics[width=0.4\textwidth]{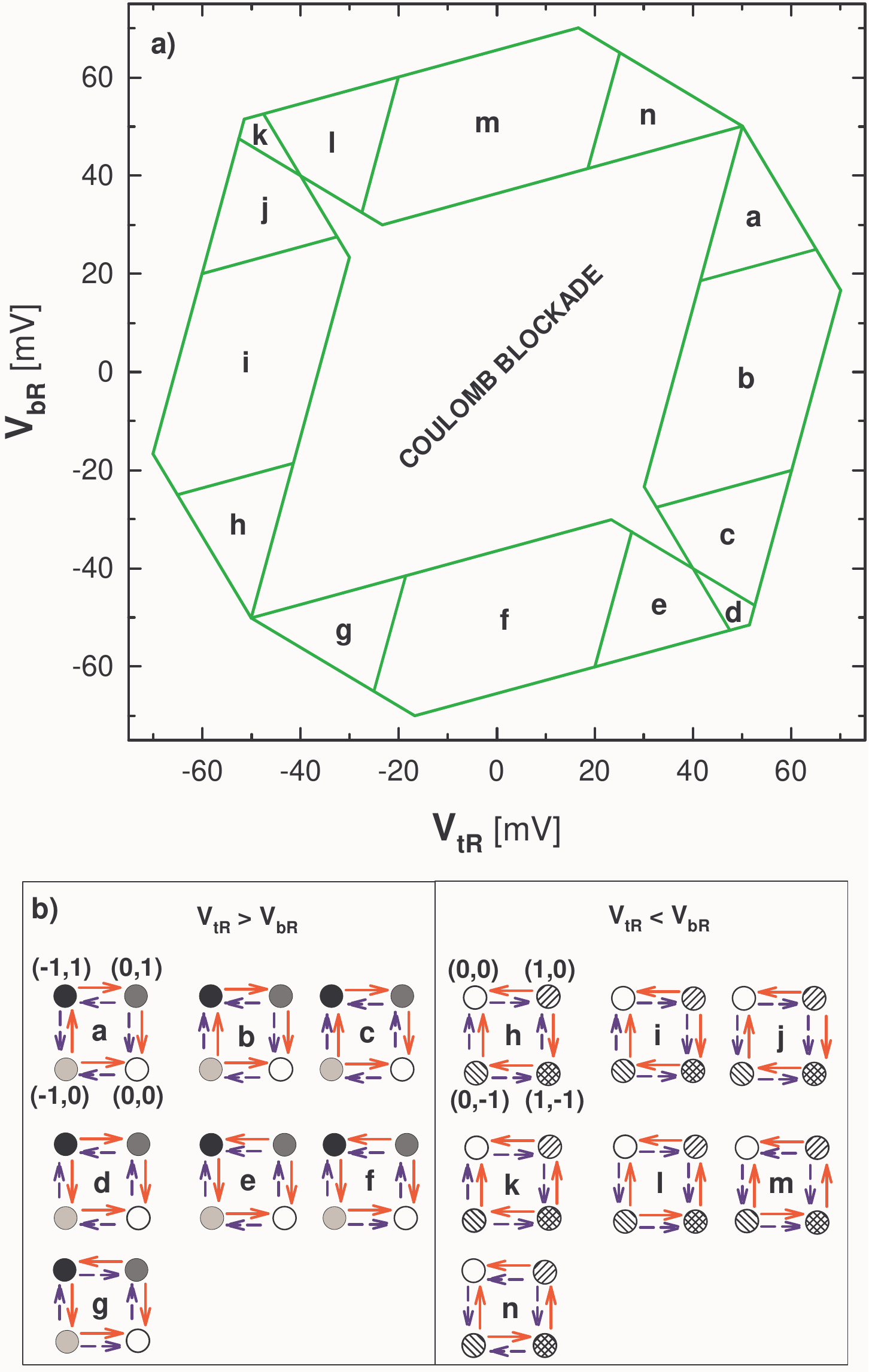}
 \caption{(Color online) (a) The diagram of the available charge states and allowed tunneling processes contributing to transport in the space of bias voltages $(V_{tR},V_{bR})$ for a strong coupling case $C_{int} = 3$ aF $> C_{tL} = C_{tR} = C_{bL} = C_{bR} = 1$ aF. (b) The diagrammatic representation of the tunneling processes $\{ \Gamma^s_{\ell\alpha}(n_t,n_b) \}$ in the charge state space in various regions $\{ a,\ldots,n \}$ (notation is the same as in Fig.~\ref{fig2}).}\label{fig3}
\end{figure}

The charging diagram and the relevant tunneling processes for the strong inter-coupling case ($C_{int}=3$ aF) are presented in Fig.~\ref{fig3}. We have found that the ground state is always $(0,0)$ in the CB region, with a global minimum of the system. However, a set of local minima appears at $(n, -n)$ (with $n=\pm 1, \pm 2, \ldots$). These polarized states $(n,-n)$ can be only reached in CB through thermally activated neighboring states $(n, -n \pm 1)$, $(n \pm 1, -n)$. Because the corresponding free energy increases with the polarization $P\equiv n_t-n_b$, thus these states are less stable for large $P$. The system evolves between the local minima in order to reach the ground state, and transition times are shorter between the metastable states with a large polarization $P$. One can expect that the polarized states and fluctuations of polarization play a relevant role in currents and shot noise. For the case $C_{int}=3$ aF the local minima appear for the polarized states $(1, -1)$ and $(-1,1)$, and all transfer rates $\Gamma^s_{\ell\alpha}(1,-1)$ and $\Gamma^s_{\ell\alpha}(-1,1)$ are exponentially small in the CB region. Comparing the diagrams in Fig.~\ref{fig2}(a) and \ref{fig3}(a) one can see that charged regions shrink with $C_{int}$. The CB region is smaller as well as charged regions $\{a,...,n\}$ above $V^{th}$. For example, in the region $b$ (at $V_{tR}>V^{th}$ and $V_{bR}\approx 0$) all four states $(0,0)$, $(-1,0)$, $(-1,1)$, $(0,1)$ participate in transport for $C_{int}=3$ aF, whereas in Fig.~\ref{fig2} only two states $(0,0)$ and $(-1,0)$ are available. For the strong inter-dot coupling the bias voltage $V_{tR}$ opens only the transfer $(0,0)\rightarrow (-1,0)$, since all other tunneling processes are allowed. One see a close loop of electron transfers between the charge states $(0,0) \rightarrow (-1,0) \rightarrow (-1,1) \rightarrow (0,1) \rightarrow (0,0)$.

\section{Analysis of currents and shot noise}

\subsection{Weak inter-dot coupling}

\begin{figure}
\includegraphics[width=0.4\textwidth]{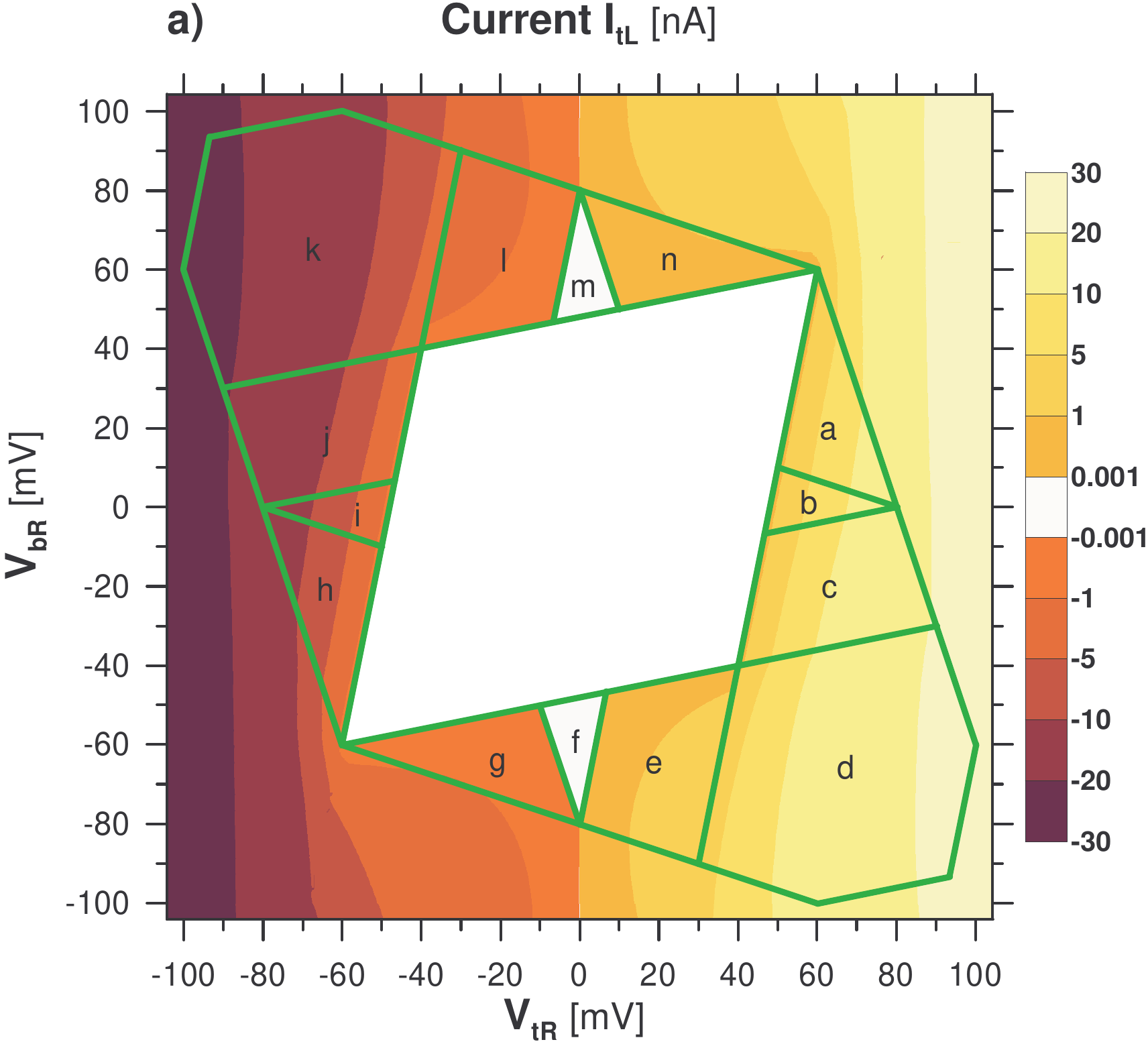} \hspace{0.5cm}
\includegraphics[width=0.4\textwidth]{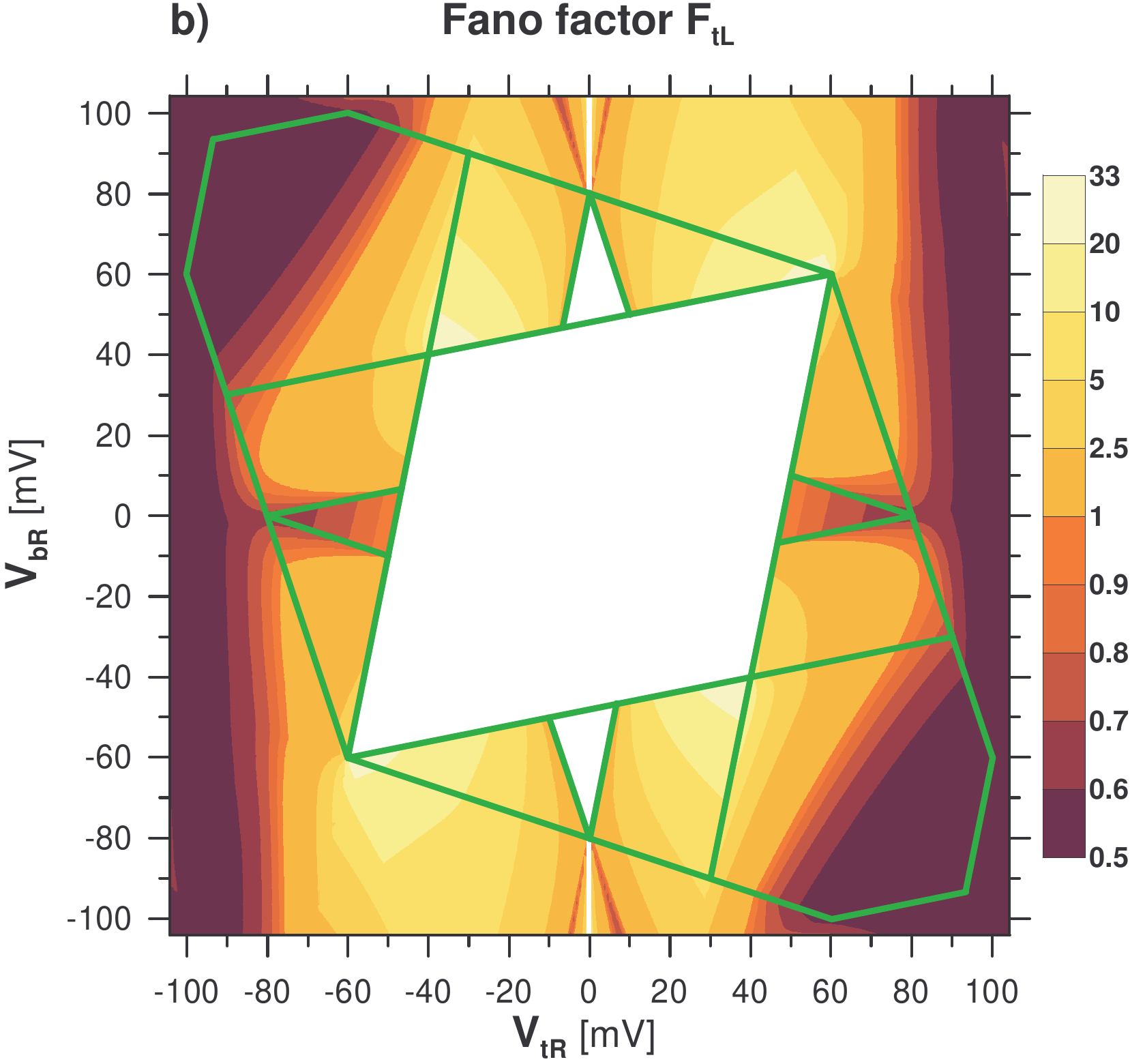} \vspace{0.5cm} \\
\includegraphics[width=0.4\textwidth]{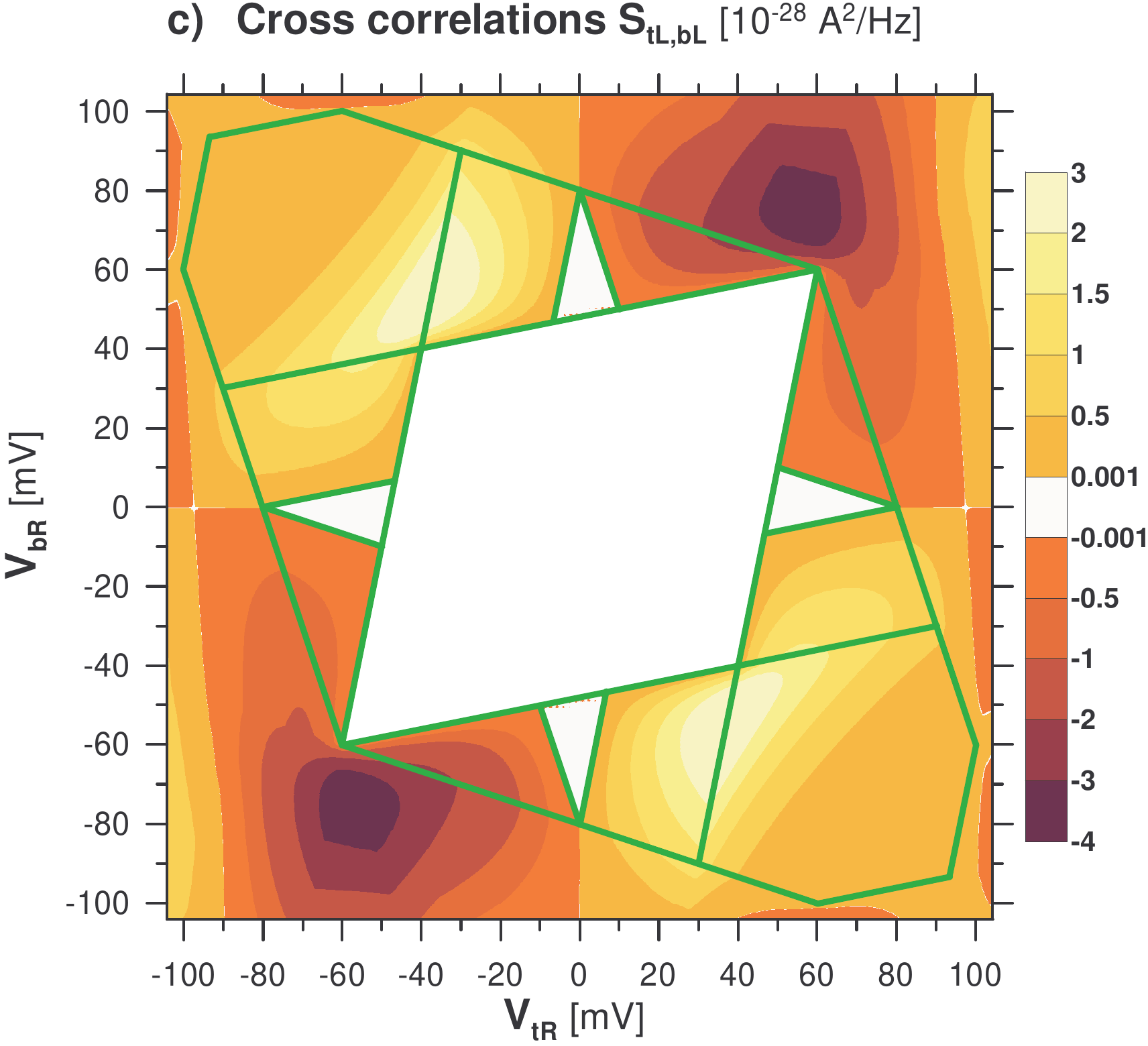} \hspace{0.5cm}
\includegraphics[width=0.4\textwidth]{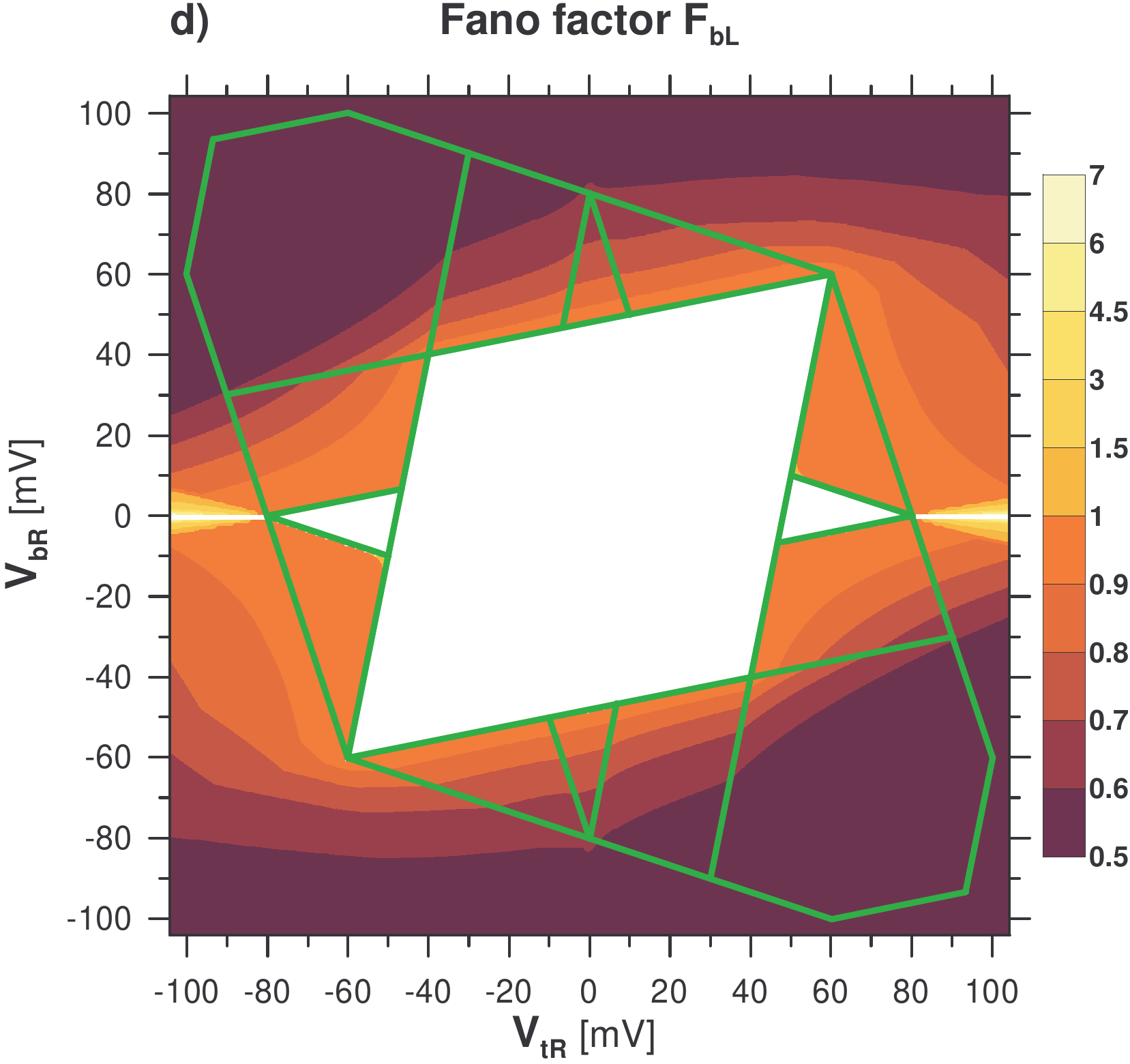}
\caption{(Color online) The current and the shot noise maps in the ($V_{tR},V_{bR})$ space for the weak inter-dot coupling $C_{int}=1$ aF. (a) The current $I_{tL}$ [nA] flowing through the top QD. (b) The Fano factor $F_{tL}$ for the left junction in the top QD. (c) The cross-correlation function $S_{tL,bL}(0)$ [$10^{-28} \mathrm{A}^2/\mathrm{Hz}$] for the currents in the top and the bottom QD. (d) The Fano factor $F_{bL}$ in the left bottom junction. The results were performed for $R_{tL} = R_{tR} = 1$ M$\Omega$, $R_{bL} = R_{bR} = 50$ M$\Omega$ and $T = 0$ K. The other parameters are the same as in Fig.~\ref{fig2}.}\label{fig4}
\end{figure}

In this section we want to study  dynamical current-current correlations. Therefore, we analyze contributions of different tunneling processes $\Gamma^s_{\ell\alpha}(n_t,n_b)$ in the $(n_t,n_b)$ charge state space to currents $I_{\ell\alpha}$ and shot noise $S_{\ell\alpha,\ell'\alpha'}$. Their voltage characteristics depend on the charging energies, on $\Delta F^s_{\ell\alpha}(n_t,n_b)$, as well as on the tunnel resistances $R_{\ell\alpha}$. Fig.~\ref{fig4} presents results of numerical calculations for the weak inter-dot coupling $C_{int}=1$ aF. We plot maps in the $(V_{tR},V_{bR})$ space for the current $I_{tL}$ through the top QD, the Fano factor $F_{tL}=S_{tL,tL}(\omega=0)/2eI_{tL}$ (for the auto-correlation current-current function in the top QD at the frequency $\omega=0$), the cross-correlation functions $S_{tL,bL}=S_{tL,bL}(\omega=0)$ between the currents in the top and the bottom QD, and the Fano factor $F_{bL}=S_{bL,bL}(\omega=0)/2eI_{bL}$ at the bottom QD. The current $I_{bL}$ is not shown, because its characteristic is very similar to $I_{tL}$ (one can rotate its map by 90 degrees). The tunnel resistances are assumed to be symmetric for the top QD ($R_{tL}=R_{tR}=1$ M$\Omega$) and for the bottom QD ($R_{bL}=R_{bR}=50$ M$\Omega$). We have analyzed the characteristics for other resistances (also for $R_{tL}=R_{tR}=R_{bL}=R_{bR}=1$ M$\Omega$), but the super-Poissonian shot noise is more pronounced in the case presented. Because temperature only smears the characteristics close to the threshold voltages $V^{th}$, so it is irrelevant for our studies, and we take $T=0$ in the calculations. We do not analyze the CB region, where other (cotunneling) processes can be relevant, and where they can dominate over contributions of sequential tunneling to currents.

Let us analyze the Fano factor $F_{tL}$ for the top QD, which is presented in Fig.~\ref{fig4}(b). In the region $b$ (for $V_{tR}>V^{th}$ and $V_{bR}\approx 0$) transport through the top QD is not disturbed by the bottom QD, and the shot noise is in the sub-Poissonian regime (with $F_{tL}<1$) as expected.  For larger positive values of $V_{bR}$ (in the region $a$, $c$, or $d$) the Fano factor $F_{tL}> 1$ and decreases below the unity with an increase of $V_{tR}$. In this case the charge states $(0,0)$, $(-1,0)$, $(-1,1)$ and $(0,1)$ participate in transport. Charge fluctuations in the bottom QD can be large and can lead to the dynamical Coulomb blockade (DCB) effect.\cite{bb00,michalek,cottetprb,cottetprl,cottetepl} The effect manifest itself in an enhancement of the shot noise to the super-Poissonian regime (with $F_{tL}>1$). It is seen in the frequency dependent part of the auto-correlation function, $S^c_{tL,tL}(\omega)$ [see Eq.~(\ref{sc})], which at $\omega=0$ is a sum of all contributions of current fluctuations with various relaxation times.\cite{bb00,michalek} This function is then positive, since many tunneling processes show bunching of electrons and their contributions are positive. Fig.~\ref{fig4}(b) shows that the super-Poissonian shot noise occurs also in the regions $e$ and $g$.

In order to understand a role of various tunneling processes in the shot noise $F_{tL}$ we analyze all components $D_{tL,tL}^{n'_tn'_b,n_tn_b}$, which contribute to the dynamical part of the function $S^c_{tL,tL}$ [see Eqs.~(\ref{sc})-(\ref{scdyn})]. The results are presented in Fig.~\ref{fig5}(a), which is a cross-section through the regions $a$, $b$, $c$ and $d$ in Fig.~\ref{fig4}(b). The auto-correlation function $D_{tL,tL}^{-10,-10}$ describes correlations of tunneling processes between the charge states $(-1,0)\rightarrow (0,0)$. In the region $b$ the function is negative, and therefore, the shot noise is sub-Poissonian. But the function increases and becomes positive in the regions $a$ and $c$. A more pronounced increase is seen for the component $D_{tL,tL}^{-11,-11}$. In contrast, the cross-correlation function $D_{tL,tL}^{-10,-11}$ is negative in all the regions. It means that the tunneling processes $(-1,0)\rightarrow (0,0)$ and $(-1,1)\rightarrow (0,1)$ are anti-bunched.  This contribution, however, does not compensate $D_{tL,tL}^{-10,-10}$ and $D_{tL,tL}^{-11,-11}$, and therefore, $F_{tL}>1$ in the regions $a$ and $c$. Our previous studies\cite{bb00,michalek} showed that the super-Poissonian shot noise is due to activation of charge fluctuations in the system. For the present model we have the same situation. Fig.~\ref{fig5}(b) presents the plots of voltage correlation functions $S_{V_t,V_t}$ and $S_{V_b,V_b}$ in the top and the bottom QD. It is seen a large enhancement of voltage and charge fluctuations in the regions $a$, $c$ and $d$ in the bottom QD (see the blue dash curve). Moreover, these plots are very similar to the auto-correlation components $D_{tL,tL}^{n_tn_b,n_tn_b}$, even small kinks at $V_{bR}=-36$ mV are similar in both the figures. Voltage fluctuations in the bottom QD are very strong, two orders of magnitude larger than those in the top QD.

The cross-correlation function $S_{tL,bL}$ is presented in Fig.~\ref{fig4}(c). We do not normalize the cross-correlation functions, because the function is a sum of two contributions proportional to some components of $I_{tL}$ and $I_{bL}$, and therefore, there is not a good normalization factor\cite{bb08} (neither $I_{tL}$, nor $I_{bL}$, nor $\sqrt{I_{tL}I_{bL}}$ can be used\cite{cottetprb,norm}). One see in Fig.~\ref{fig4}(c) that the function $S_{tL,bL}$ is negative in the right-upper and the left-lower quarter of the map, when the currents $I_{tL}$ and $I_{bL}$ have the same direction. This dependence is typical for cross-correlation in scattering of fermions, which exhibit anti-bunching.\cite{feynman,kulik,khlus,lesovik,buttiker1990,blanter,bb08} However, in our situation electronic transfer is separated in the top and in the bottom part of the device, so anti-bunching is not fermionic nature. We show later that correlation of both the electronic channels is due to Coulombic interaction and potential fluctuations.

In contrast to $F_{tL}$, the current noise $F_{bL}$ in the bottom QD, is generally the sub-Poissonian type, except with a narrow stripe around $V_{bR} = 0$ for high $V_{tR}$ [see Fig.~\ref{fig4}(d)]. The difference between the maps in Fig.~\ref{fig4}(b) and \ref{fig4}(d) results from the DCB effect. An electron traversing the bottom channel spends a long time at the bottom QD and the charge fluctuations at the top QD are relatively fast. Therefore, transport in the bottom channels depends on an average potential in the top QD and the shot noise $F_{bL}$ is very little sensitive to individual tunneling events in the top channel of our device.

\subsection{Strong inter-dot coupling}

Maps in Fig.~\ref{fig6} show transport characteristics $I_{tL}$, $F_{tL}$, $S_{tL,bL}$ and $F_{bL}$ in the bias space $(V_{tR},V_{bR})$ for the strong inter-dot coupling. In this case the maps are different than for the weak coupling in Fig.~\ref{fig4}, because the charging diagram is different (compare Figs.~\ref{fig3}~and~\ref{fig2}). In the regions $\{a,...,g\}$ all four charge states $(0,0)$, $(-1,0)$, $(-1,1)$, $(0,1)$, contribute to transport;  whereas the states $(0,0)$, $(1,0)$, $(1,-1)$, $(0,-1)$ are active in the regions $\{h,...,n\}$. Tunneling processes and their contributions to transport are different than in the case of the weak coupling. Let us analyze the Fano factor $F_{tL}$ presented in Fig.~\ref{fig6}(b). In the region $b$ (for $V_{tR}>V^{th}$ and $V_{bR}\approx 0$) the shot noise is super-Poissonian, but one see  also a dark area (with $F_{tL}<1$) inside this region.  It means that there are various competing tunneling processes and their role changes with the bias voltage. An interesting situation is in the region $f$ (for $V_{bR}>V^{th}$ and small $V_{tR}$). In the middle $F_{tL}$ shows a peak, while in Fig.~\ref{fig4}(b) the shot noise decreased below the unity. One see the dark areas in Fig.~\ref{fig6}(b) close to the borders with the regions $e$ and $g$.

We have analyzed several cross-sections of the maps in Fig.~\ref{fig6}. One of them is in Fig.~\ref{fig7}, which presents $F_{tL}$  in the region $b$ at $V_{bR} = 0$. We also plotted the dynamical part of the shot noise, $F^c_{tL}=S^c_{tL,tL}/2eI_{tL}$, and its components $D_{tL,tL}^{n'_tn'_b,n_tn_b}/2eI_{tL}$.  For small voltages $V_{bR}$ the auto-correlation components $D_{tL,tL}^{-10,-10}$ and $D_{tL,tL}^{-11,-11}$ dominate over the cross-correlation one $D_{tL,tL}^{-10,-11}$, so $S^c_{tL,tL} > 0$ and the corresponding $F_{tL}$ is super-Poissonian. It is seen that the function $D_{tL,tL}^{-10,-11}$ always decreases with $V_{tR}$. The auto-correlation components have different dependencies, and these two components compete each other. The function $D_{tL,tL}^{-10,-10}$ increases, whereas $D_{tL,tL}^{-11,-11}$ decreases. For small bias voltage tunneling processes $(-1,1)\rightarrow (0,1)$ dominates in the super-Poissonian shot noise, while auto-correlations $(-1,0)\rightarrow (0,0)$ become dominating for large $V_{tR}$. In the intermediate voltage range $F_{tL}<1$, because the cross-correlation component plays a dominating role.

In the Fig.~\ref{fig8} we present $F_{tL}$ and its relevant components at $V_{bR} = -50$ mV for a cross-section through the regions $f$, $e$ and $d$. The strong enhancement of the $F_{tL}$ in the region $f$ at small $V_{tR}$ is due to the Schottky term $S^{Sch}_{tL}$. This situation is in contrast to that one for $V_{bR}=0$ presented in Fig.~\ref{fig7}, where $S^{Sch}_{tL}/2eI_{tL}=1$ in whole region $b$ (with exception of a small region close to the threshold voltages, where strong thermal current fluctuations occur). From the plots in Fig.~\ref{fig8} one see that the contribution $S^{Sch}_{tL}$ to the total Fano factor decreases with increasing voltage $V_{tR}$, while a role of the dynamical part $S^c_{tL}$ grows. This leads to the reduction of the Fano factor below Poissonian value for voltages $14.5$ mV $\lesssim V_{tR} \lesssim 24$ mV (close to the border between the region $f$ and $e$). For higher voltages $S^c_{tL,tL}$ become positive and the Fano factor is super-Poissonian. One see from Fig.~\ref{fig8} that an enhancement of the $F^c_{tl}$ in the region $e$ is caused by auto-correlation processes $D_{tL,tL}^{-10,-10}$ (red dot curve). The border between regions $f$ and $e$ is a transition line between the sub- and the super-Poissonian shot noise, where the tunneling process $\Gamma^+_{tR}(-1,0)$ disappears while the process $\Gamma^-_{tR}(0,0)$ becomes activated.

The cross-correlation $S_{tL,bR}$ [see Fig.~\ref{fig6}(c)] is similar to that one in the weak coupling case, which shows anti-bunching. However, this anti-bunching process is not fermionic origin, exchange of electrons is impossible between the both QDs. In our system strong Coulombic repulsion occurs between charges transferred through the top and the bottom part of the device. This process mimics the Pauli exclusion principle, and therefore, electronic transfers through the top and the bottom QD should be anti-correlated and the function $S_{tL,bR}<0$. We make a cross-section through regions $a$, $b$ and $c$ at $V_{tR}=50$ mV for the current $I_{bL}$ and $S_{tL,bR}$ [see Fig.~\ref{fig9}(a)]. We expected that the cross-correlation function $S_{tL,bR}$ is proportional to the current $-I_{bL}$, at least in the region $b$ for small bias. However, the plot is different -- its shows a plateau around $V_{bR}\approx 0$ and a sharp increase, which is proportional to $-I_{bL}$. It is clear that the current $I_{bL}$ and the cross-correlation function $S_{tL,bR}$ may change  their dependencies in the regions $a$ and $c$, because new tunneling processes are activated. In order to understand the dependence of $S_{tL,bR}$ in the region $b$ we plot its components $D_{tL,bL}^{n'_tn'_b,n_tn_b}$ in Fig.~\ref{fig9}(b). Notice that at $V_{bR}=0$ all the components are non-zero, but they compensate each other and the sum is $S_{tL,bR}=0$. It is clear that $D_{tL,bL}^{-10,-10}+D_{tL,bL}^{-10,01}=0$ at $V_{bR}=0$, because these components describe correlations between the tunneling event $(-1,0)\rightarrow (0,0)$ through the left top junction and the forward and the backward tunneling event through the left bottom junction (between the charge states $(-1,0)\rightarrow (-1,1)$ and $(0,1)\rightarrow (0,0)$, respectively). Similarly, one has for $D_{tL,bL}^{-11,-10}+D_{tL,bL}^{-11,01}=0$. However, one can see that at $V_{bR}=-3.55$ mV the components $D_{tL,bL}^{-10,-10}=0$ and $D_{tL,bL}^{-10,01}=0$, while $D_{tL,bL}^{-11,-10}=0$ and $D_{tL,bL}^{-11,01}=0$ at $V_{bR}=-3.73$ mV. This compensation is due to advanced and retarded correlations, which occur for any component $D_{tL,bL}^{n'_tn'_b,n_tn_b}$ [see Eq.~(\ref{scdyn}) and terms corresponding to $\omega$ and $-\omega$].\cite{symmetric} Both the contributions, advanced and retarded, compensate each other either at $V_{bR}=-3.55$ mV or -3.73 mV, respectively. This effect is also responsible for the plateau seen in $S_{tL,bR}$. Outside the plateau region the dominating components are $D_{tL,bL}^{-10,01}$ and $D_{tL,bL}^{-11,-10}$ (for negative and positive $V_{bR}$, respectively) -- see the pink long dash and the blue short dash curves in the bottom part of Fig.~\ref{fig9}.

Let us come back to the maps in Fig.~\ref{fig6} and analyze the Fano factor $F_{bL}$ in the bottom QD shown in Fig.~\ref{fig6}(d). It is the sub-Poissonian type in the most regions, but the function has an interesting feature in the region $b$ and $i$, where $F_{bL}>1$. The shot noise is due to charge fluctuations in the top QD. When an electron leaves the top QD, $(0,0)\rightarrow (-1,0)$, the system becomes unstable, therefore, an electron is attracted to the bottom QD, $(-1,0)\rightarrow (-1,1)$, through the left or the right tunnel junction. The system is in the metastable state $(-1,1)$. The next tunneling event in the top QD, $(-1,1)\rightarrow(0,1)$, leads to pushing out the electron from the bottom QD, $(0,1)\rightarrow (0,0)$. This is a pumping effect -- a charge injected into (or ejected from) the top QD leads to ejection (injection) of a charge at the bottom QD. One can expect that the shot noise $S_{bL,bL}$ should be proportional to the tunneling current $I_{tL}$. It is easily to prove it, calculating these quantities in the region $b$ for $V_{bR}=0$. The results are presented in Fig.~\ref{fig10}. The main contribution to the shot noise [Eq.~\ref{noise}] is due to the Schottky term [Eq.~\ref{sch}], which is now $S^{Sch}_{bL}=4eI^+_{bL}$. Since for $V_{bR}=0$ the current flowing into and from the bottom QD are equal $I^+_{bL}=I^-_{bL}$, the total current $I_{bL}=I^+_{bL}-I^-_{bL}=0$. Therefore, Fig.~\ref{fig10} presents the auto-correlation function $S_{bL,bL}$, instead the Fano factor $F_{bL}=S_{bL,bL}/2eI_{bL}$. In this case one can obtain analytical formula
\begin{eqnarray}\label{Ilb}
I^+_{bL}=e\frac{\Gamma^-_{tR}(0,0)\Gamma^+_{tL}(-1,1)\Gamma^+_{b}(-1,0)\Gamma^-_{b}(0,1)}{2w}\;,\\
\end{eqnarray}
where $w=\Gamma^-_{tR}(0,0) \{\Gamma^-_{b}(0,1)[\Gamma^+_{b}(-1,0)+\Gamma^+_{tL}(-1,1)]+
\Gamma^+_{b}(-1,0)[\Gamma^+_{tL}(-1,1)+\Gamma^-_{tR}(0,1)]\}+
\Gamma^-_{b}(0,1)[\Gamma^+_{b}(-1,0)+\Gamma^+_{tL}(-1,0)]\Gamma^+_{tL}(-1,1)$. Here, we used $R_{bL}=R_{bR}$ and denoted $\Gamma^+_{b}(-1,0)=2\Gamma^+_{bL}(-1,0)$, $\Gamma^-_{b}(0,1)=2\Gamma^-_{bL}(0,1)$.
 The auto-correlation function is derived from Eq.~(\ref{noise})
\begin{eqnarray}\label{slb}
S_{bL,bL}(0)=e^2 \frac{\Gamma^-_{tR}(0,0)\Gamma^+_{tL}(-1,1)\Gamma^+_{b}(-1,0) \Gamma^-_{b}(0,1) }{w}\;.
\end{eqnarray}
In this case one see that the Schottky term $S^{Sch}_{bL}=2 S_{bL,bL}(0)$. The current through the top QD is
\begin{eqnarray}
I_{tL}=e\frac{\Gamma^-_{tR}(0,0) \Gamma^+_{tL}(-1,1)\{\Gamma^-_{tR}(0,1)\Gamma^+_{b}(-1,0)+
\Gamma^-_{b}(0,1)[\Gamma^+_{tL}(-1,0)+\Gamma^+_{b}(-1,0)]\}}{w}\;.
\end{eqnarray}
The tunneling process between the states $(0,0)\rightarrow (-1,0)$, described by $\Gamma^-_{tR}(0,0)$, is relevant for $I_{tL}$ as well as $S_{bL,bL}$. Therefore, the shot noise in the bottom QD (see Fig.~\ref{fig10}) is proportional to the current in the top QD.

\subsection{Asymmetric resistances}

So far we have analyzed the system in the case when each QD is symmetrically coupled to the left and the right electrodes. In this section we will show  how asymmetrical coupling to the leads can influence noises in the system. Let us analyze first the weak coupling case ($C_{int}=1$ aF). The maps for the Fano factor $F_{tL}$ and the cross-correlation function $S_{tL,bL}$ are presented in Fig.~\ref{fig11}. The value of the Fano factor $F_{tL}$ is slightly smaller than in the symmetrical case [compare with Fig.~\ref{fig4}(b)]. One can see that now the sub-Poissonian shot noise dominates in the region $a$. The factor $F_{tL}$ has non-monotonic behavior, which is well seen in the regions $a$, $c$ and $d$. Its value is super-Poissonian in the area very close to the CB and close to the region $e$, next it drops below the unity with an increase of $V_{tR}$ and  grows to the super-Poissonian values for higher voltages. The super-Poissonian shot noise occurs also in the regions $e$ and $g$. We have found that, as in the symmetrical case, for an enhancement of $F_{tL}$ the auto-correlation processes are responsible, while the cross processes lead to a reduction of $F_{tL}$ below unity. The cross-correlation function $S_{tL,bL}$ is presented in Fig.~\ref{fig11}(b). In contrast to the symmetrical case [Fig.~\ref{fig4}(c)], when $S_{tL,bL}$ is always negative in the right-upper and the left-lower quarter of the map, now one see that it can be also positive in some areas. This dependence is non typical for cross-correlation in scattering of electrons and now it exhibits bunching.

The results for the strong coupling case ($C_{int}=3$ aF) are presented in Fig.~\ref{fig12}. One see that behavior of $F_{tL}$ is very similar to that one obtained in the symmetrical case [compare Figs.~\ref{fig6}(b)~and~\ref{fig12}(a)]. The cross-correlation function $S_{tL,bL}$ in Fig.~\ref{fig12}(b) is non-monotonic, it can even change its sign with an increase of the bias voltage (see, for example, the region $a$ and $b$, where $S_{tL,bL}<0$ in the upper quarter and becomes positive for higher $V_{tR}$). We made a cross-section for $V_{tR}=40$ mV, through regions  $b$ and $c$, in order to analyze different contributions $D_{tL,bL}^{n'_tn'_b,n_tn_b}$ to $S_{tL,bL}$ -- see Fig.~\ref{fig13}. For negative $V_{bR}$ the component $D_{tL,bL}^{-10,01}$ is dominating (see the pink long dash curve). This contribution is also responsible for change of $S_{tL,bL}$ from positive to negative values with increasing $V_{bR}$. When $V_{bR}>0$ the amplitude of the component $D_{tL,bL}^{-11,-10}$ and $D_{tL,bL}^{-11,01}$ is larger than the other components. Moreover, $D_{tL,bL}^{-11,-10}$, describing correlations between tunneling events through the left top junction from the charge states $(-1,1)$ and forward tunneling processes in the bottom QD, becomes to play a dominating role.

\section{Summary and final remarks}

\begin{figure}\includegraphics[width=0.4\textwidth]{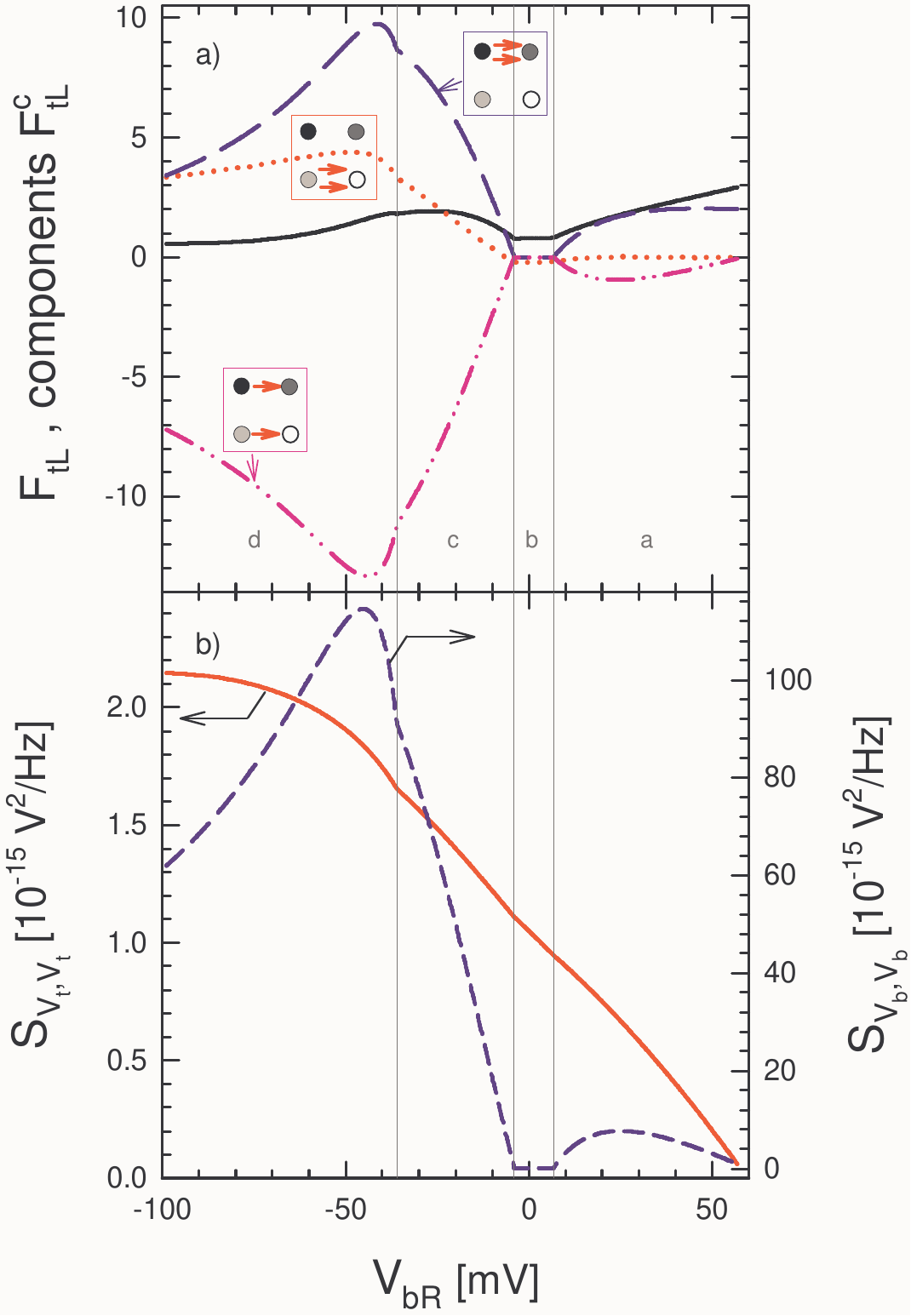}
\caption{(Color online) (a) Plots of the Fano factor $F_{tL}$ (black solid curve) and all its components $D_{tL,tL}^{n'_tn'_b,n_tn_b}$ contributing to the dynamical part of the shot noise $S^c_{tL,tL}$ for $V_{tR}=$ 60 mV [the cross-section of the map in Fig.~\ref{fig4}(b) through regions $a$, $b$, $c$ and $d$]. The red dotted curve is for $D_{tL,tL}^{-10,-10}$, which corresponds for the auto-correlation function of the tunneling processes $(-1,0)\rightarrow (0,0)$. Similarly, the blue long dash curve is for $D_{tL,tL}^{-11,-11}$. The pink dash-dot-dot curve corresponds to $D_{tL,tL}^{-10,-11}$, which describes the cross-correlation of the tunneling events $(-1,0)\rightarrow (0,0)$ and $(-1,1)\rightarrow (0,1)$. All the curves are normalized as the Fano factor (i.e., they are divided by $2e I_{tL}$). (b) Plots of the voltage-voltage correlation function $S_{V_t,V_t}$ (red solid curve) and $S_{V_b,V_b}$ (blue short dash curve) describing potential and charge fluctuations at the top and the bottom QD, respectively. Note that the right hand axis corresponds to  $S_{V_b,V_b}$ and potential fluctuations at the bottom QD are two order of magnitude larger than those in the top QD.}\label{fig5}
\end{figure}

We performed studies of dynamical current correlations in the system of two capacitively coupled quantum dots (2QDs). In this 4-terminal system we first determined stability diagrams for states with various number of additional charge introduced either to the top or the bottom QD as a function of bias voltage $V_{tR}$ and $V_{bR}$ applied to the top and the bottom channel, respectively. The charge diagram is a background for studies current-current correlations, because we can separate allowed tunneling processes, which contribute to currents. We considered two different situations for weak and strong coupling between QDs. For weak coupling one can observe independent transport through one of QD in some range of bias voltages $V_{tR}$ and $V_{bR}$ (as one can expect). In contrast, for the strong coupling case, where both QDs are engaged in transport, the charging diagram is different. In the Coulomb blockade region the charge states $(-1,1)$ or $(1,-1)$ are metastable and when bias is larger than the threshold voltage they participate in transport [together with the state $(0,0)$ and its neighbors]. For a very strong coupling a larger charge space with the states $(-n,n)$ and $(n,-n)$ (for $n>1$) can be engaged. Transport through one of the channel induces large potential fluctuations in the second one [see, e.g., $S_{V_t,V_t}$ and $S_{V_b,V_b}$ in Fig.~\ref{fig5}(b)]. This is a dynamical process and it is responsible for large enhancement of the Fano factor. We called it the \textit{dynamical Coulomb blockade effect}.\cite{bb00,michalek} We decomposed the correlation function $S_{\ell\alpha,\ell'\alpha'}$ into individual contributions $D_{\ell\alpha,\ell'\alpha'}^{n'_tn'_b,n_tn_b}$ of tunneling processes, which occur in the charge space $(n_t,n_b)$. Dynamical contributions are relevant for the Fano factor. Their auto-correlated components are positive, whereas the cross-correlated components are negative. They compensate  each other in part, and lead to the super- or the sub-Poissonian shot noise, depending on which contribution dominates (either auto-correlation or cross-correlation processes).

\begin{figure}
\includegraphics[width=0.4\textwidth]{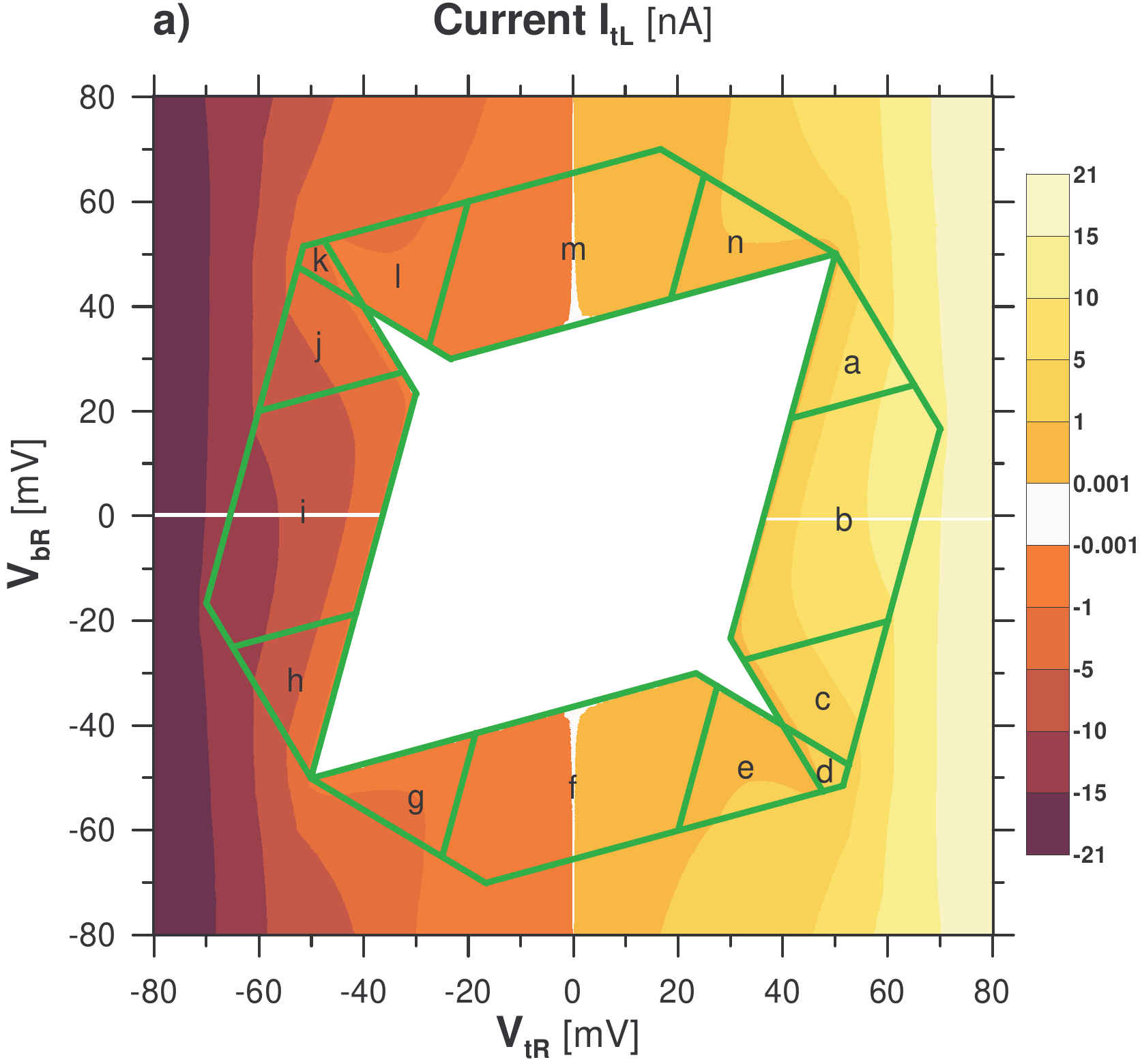} \hspace{0.5cm}
\includegraphics[width=0.4\textwidth]{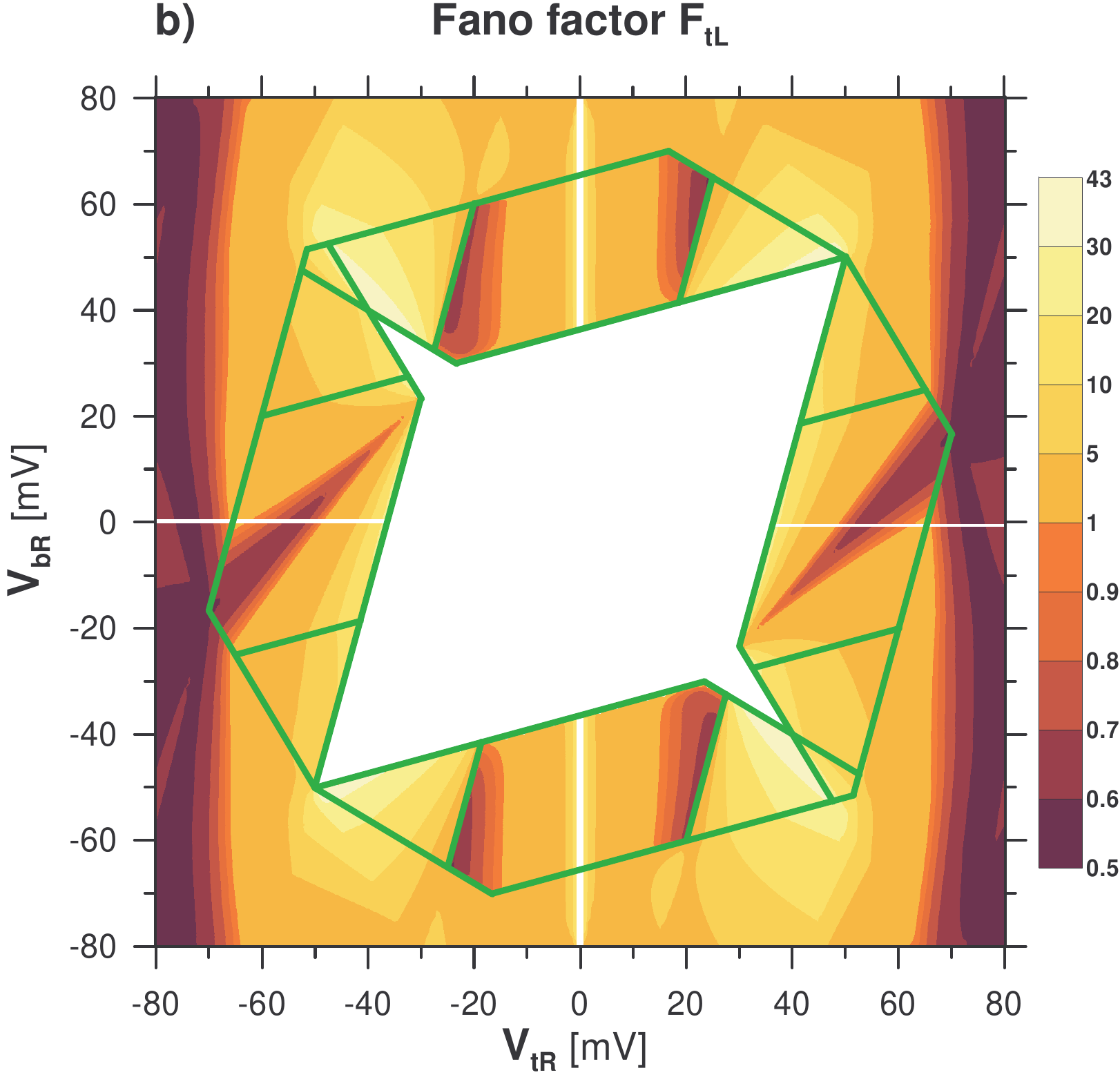} \vspace{0.5cm} \\
\includegraphics[width=0.4\textwidth]{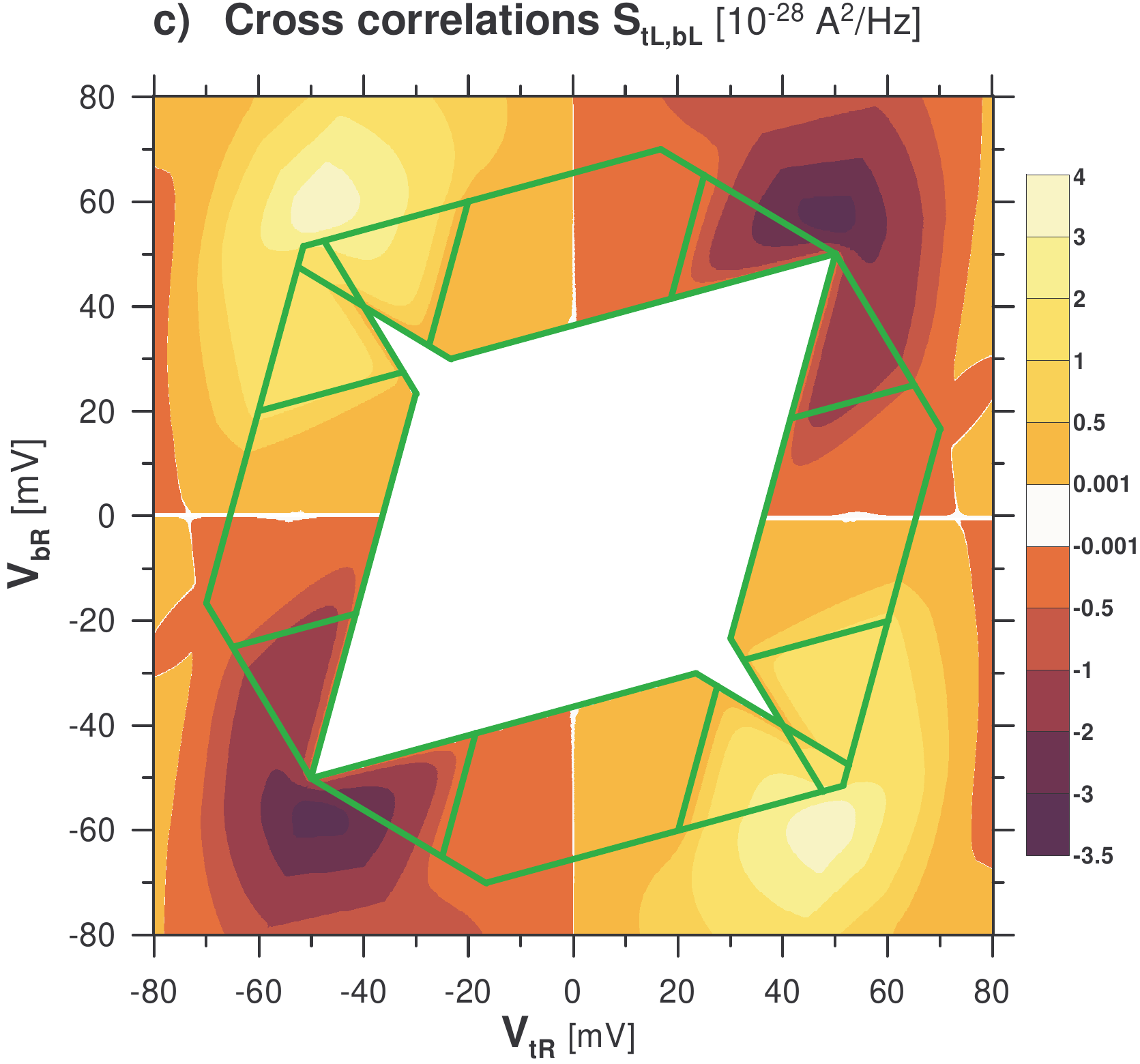} \hspace{0.5cm}
\includegraphics[width=0.4\textwidth]{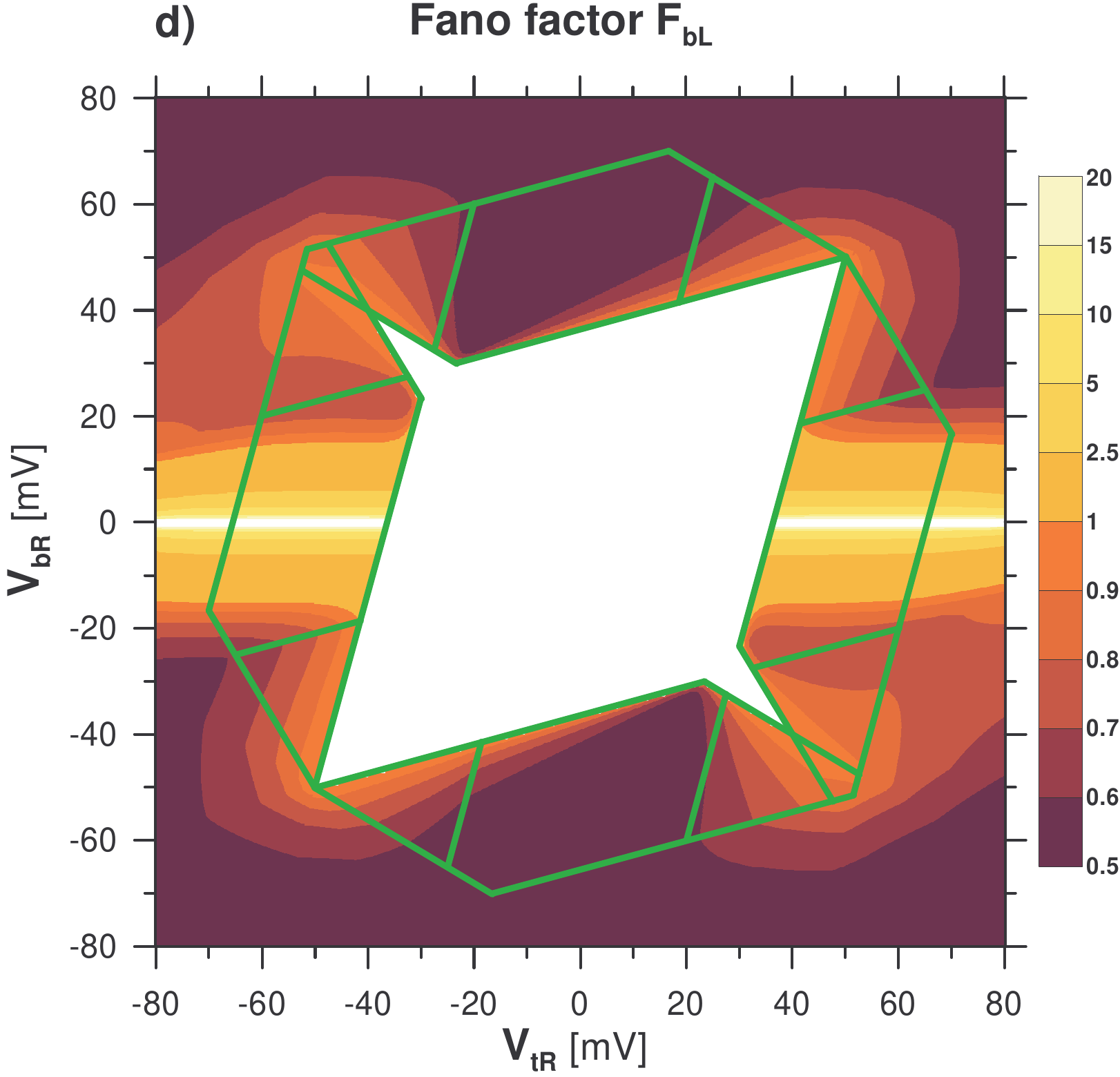}
\caption{(Color online) The current and the shot noise maps for the strong inter-dot coupling $C_{int}=3$ aF. (a) The current $I_{tL}$ [nA] flowing through the top QD. (b) The Fano factor $F_{tL}$ for the left junction in the top QD. (c) The cross-correlation function $S_{tL,bL}(0)$ [$10^{-28} \mathrm{A}^2/\mathrm{Hz}$] for the currents in the top and the bottom QD. (d) The Fano factor $F_{bL}$ in the left bottom junction. The results were performed for $R_{tL} = R_{tR} = 1$ M$\Omega$, $R_{bL} = R_{bR} = 50$ M$\Omega$ and $T = 0$ K. The other parameters are the same as in Fig.~\ref{fig3}.}\label{fig6}
\end{figure}

\begin{figure}
\includegraphics[width=0.4\textwidth]{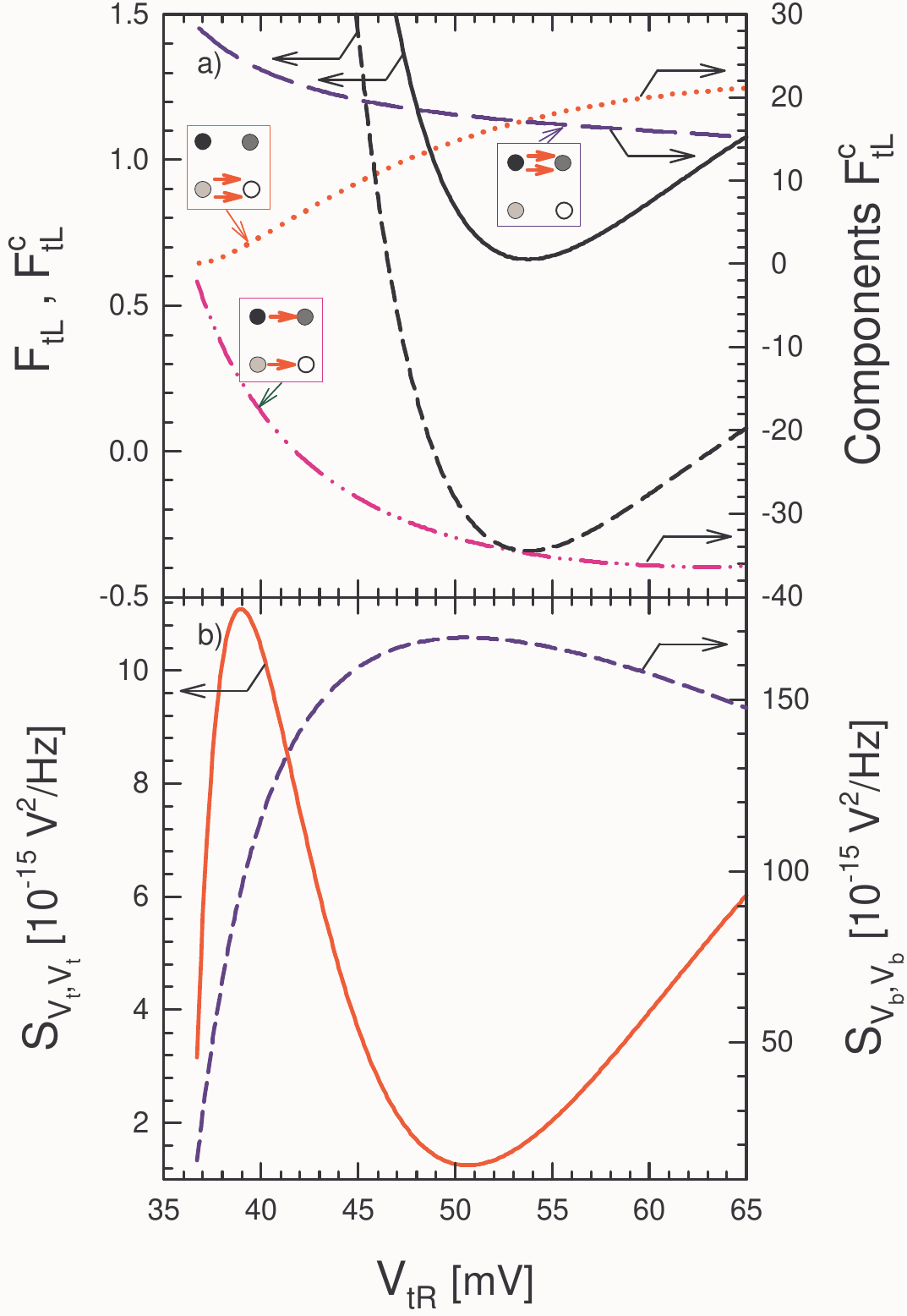}
\caption{(Color online) (a) The Fano factor $F_{tL}$ (black solid line), $F^c_{tL}$ (black short dash line) and its relevant components: $D_{tL,tL}^{-10,-10}$ (red dotted curve), $D_{tL,tL}^{-11,-11}$ (blue long dash curve), and $D_{tL,tL}^{-10,-11}$ (pink dash-dot-dot curve) plotted at $V_{bR}=0$ mV. All the components are normalized as the Fano factor. The other parameters are the same as in Fig.~\ref{fig6}. (b) Voltage correlation functions $S_{V_t,V_t}$ (red solid curve) and $S_{V_b,V_b}$ (blue short dash curve) in the top and the bottom QD, respectively.}\label{fig7}
\end{figure}

\begin{figure}
\includegraphics[width=0.4\textwidth]{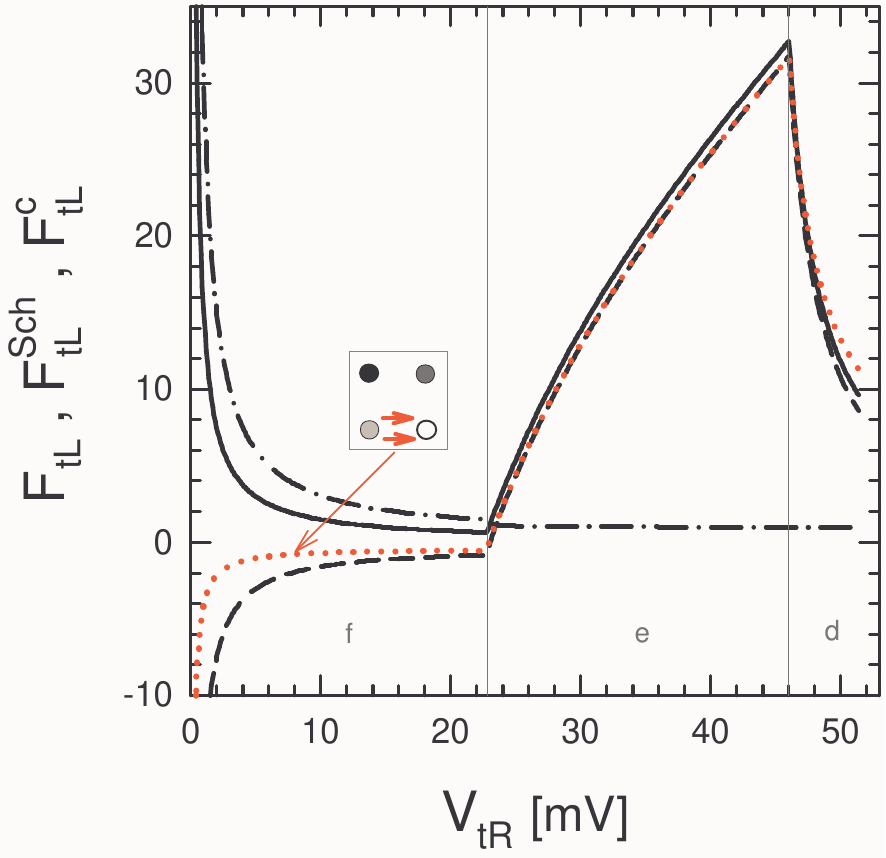}
\caption{(Color online) The Fano factor $F_{tL}$ (black solid line) and its components: the Schottky term $F^{Sch}_{tL}$ (black dash-dot curve) and the dynamical part $F^c_{tL}$ (black short dash curve) for $V_{bR}= -50$ mV (a cross-section through the regions $f$, $e$, and $d$). The red dotted curve represents a component $D_{tL,tL}^{-10,-10}$, which is a relevant contribution to the dynamical part of the shot noise. The component $D_{tL,tL}^{-10,-10}$ is normalized as the Fano factor. In the region $e$ the plot $D_{tL,tL}^{-10,-10}$ is very close to $F^c_{tL}$. The other parameters are the same as in Fig.~\ref{fig6}.}\label{fig8}
\end{figure}

\begin{figure}
\includegraphics[width=0.4\textwidth]{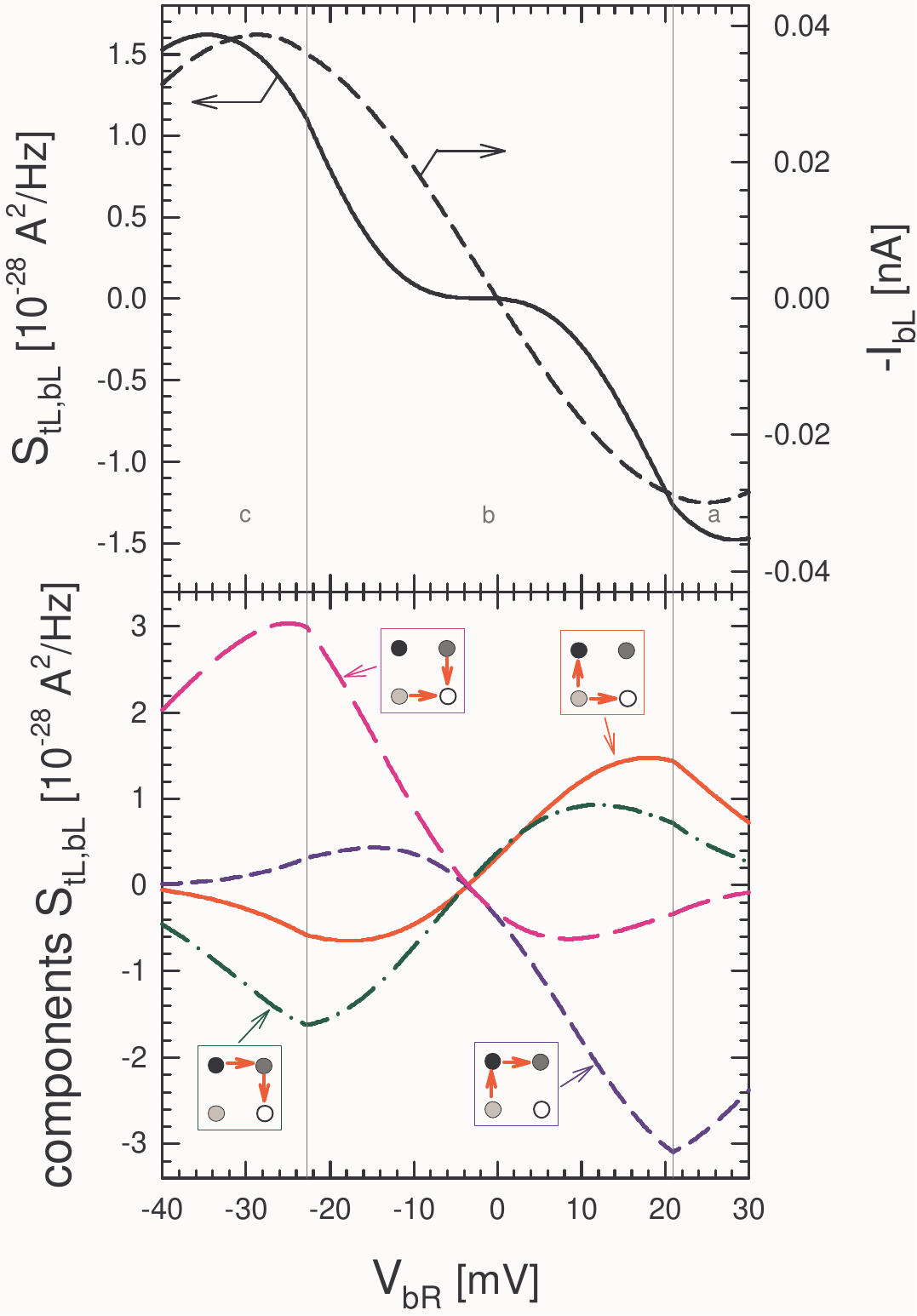}
\caption{(Color online) (a) The cross-correlation function $S_{tL,bL}$ (black solid curve) and the current $I_{bL}$ (black short dash curve) in the bottom QD plotted vs. $V_{bR}$ for the cross-section through regions $a$, $b$ and $c$ at $V_{tR}=50$ mV. The other parameters are the same as in Fig.~\ref{fig6}. (b) The components of $S_{tL,bL}$: the pink long dash curve corresponds to $D_{tL,bL}^{-10,01}$, which describes correlations of the tunneling events $(-1,0)\rightarrow (0,0)$ through the left top tunnel barrier and $(0,1)\rightarrow (0,0)$ through the left bottom tunnel barrier; the red solid curve corresponds to $D_{tL,bL}^{-10,-10}$, which describes correlations between $(-1,0)\rightarrow (0,0)$ and $(-1,0)\rightarrow (-1,1)$; the green dash-dot curve corresponds to $D_{tL,bL}^{-11,01}$, which describes correlations between $(-1,1)\rightarrow (0,1)$ and $(0,1)\rightarrow (0,0)$; and the blue short dash curve corresponds to $D_{tL,bL}^{-11,-10}$, which describes correlations between $(-1,1)\rightarrow (0,1)$ and $(-1,0)\rightarrow (-1,1)$.}\label{fig9}
\end{figure}

\begin{figure}
\includegraphics[width=0.4\textwidth]{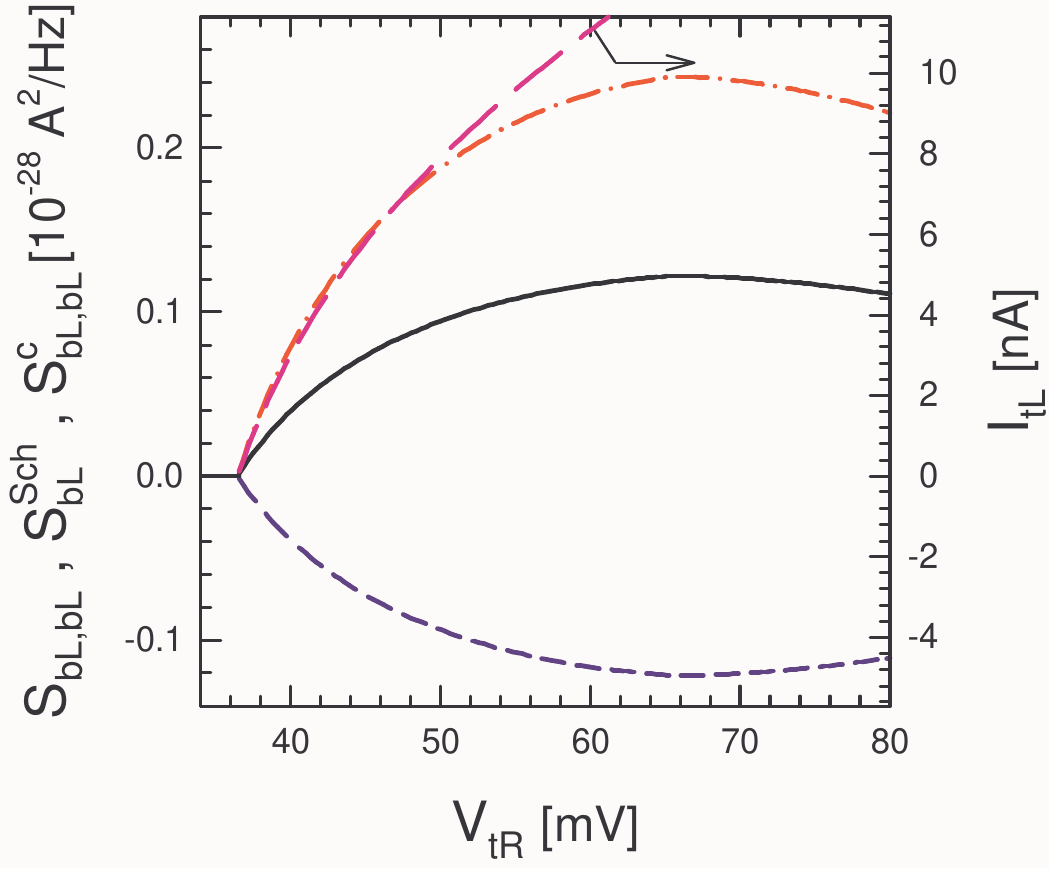}
\caption{(Color online) The function $S_{bL,bL}(0)$ (solid black curve) plotted vs. the bias $V_{tR}$ for $V_{bR}=0$ and $C_{int}=3$. It is a cross-section through the region $b$ of the map in Fig.~\ref{fig6}(d). We also plotted its Schottky component $S^{Sch}_{bL}$ (red dash-dot curve) and the frequency dependent part $S^c_{bL,bL}(0)$ (blue short dash curve). The plot shows that the shot noise in the bottom QD is induced by the current in the top QD $I_{tL}$ (pink long dash curve).}\label{fig10}
\end{figure}

\begin{figure}
\includegraphics[width=0.4\textwidth]{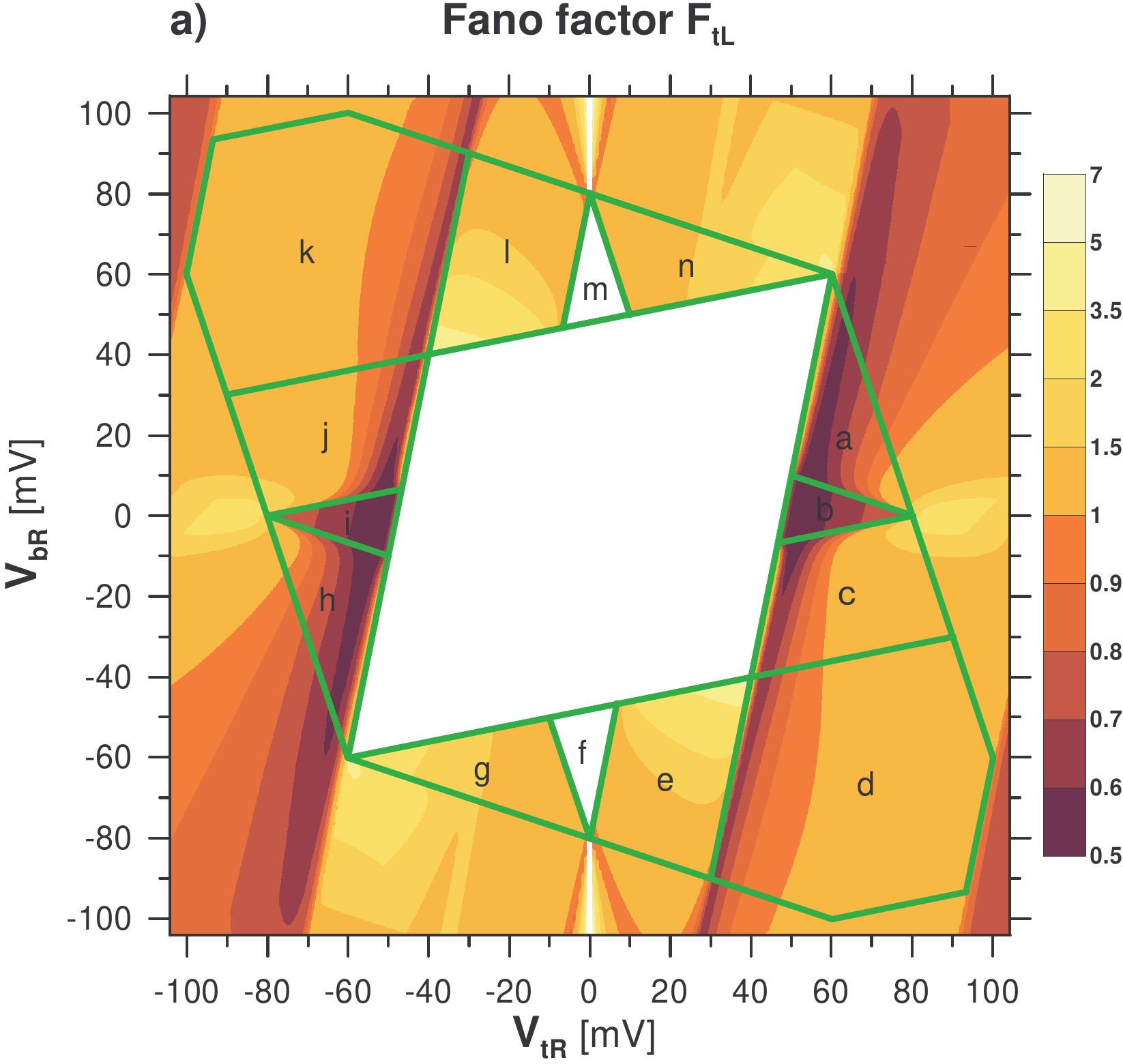}
\includegraphics[width=0.4\textwidth]{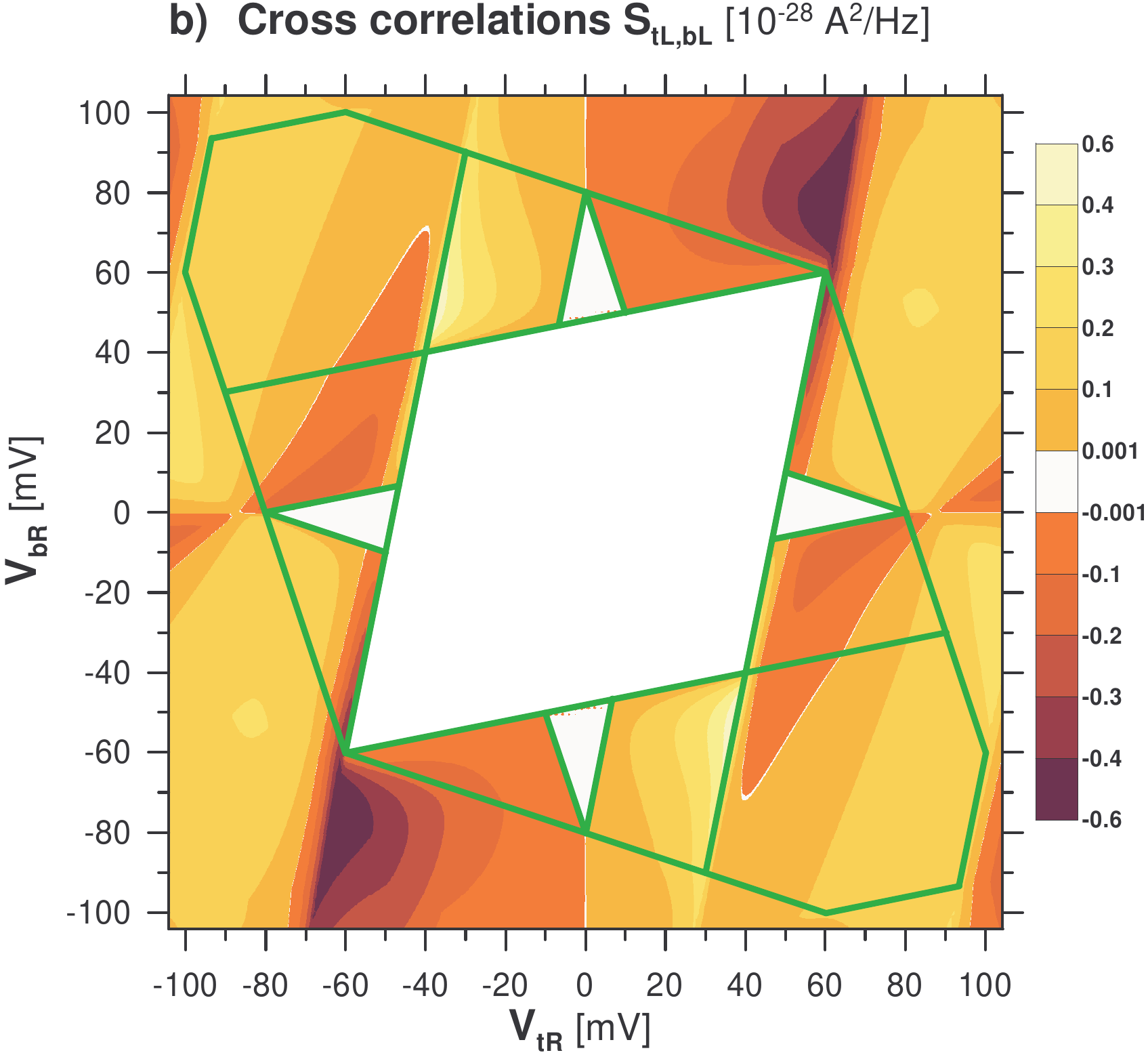}
\caption{(Color online) Maps (a) for the Fano factor $F_{tL}$ and (b) for the cross-correlation $S_{tL,bL}$ at the weak inter-dot coupling $C_{int} =1$ aF and for $R_{tL}=20$ M$\Omega$, $R_{tR}=1$ M$\Omega$, $R_{bL}=R_{bR}=50$ M$\Omega$. The other parameters are the same as in Fig.~\ref{fig2}.}\label{fig11}
\end{figure}

\begin{figure}
\includegraphics[width=0.4\textwidth]{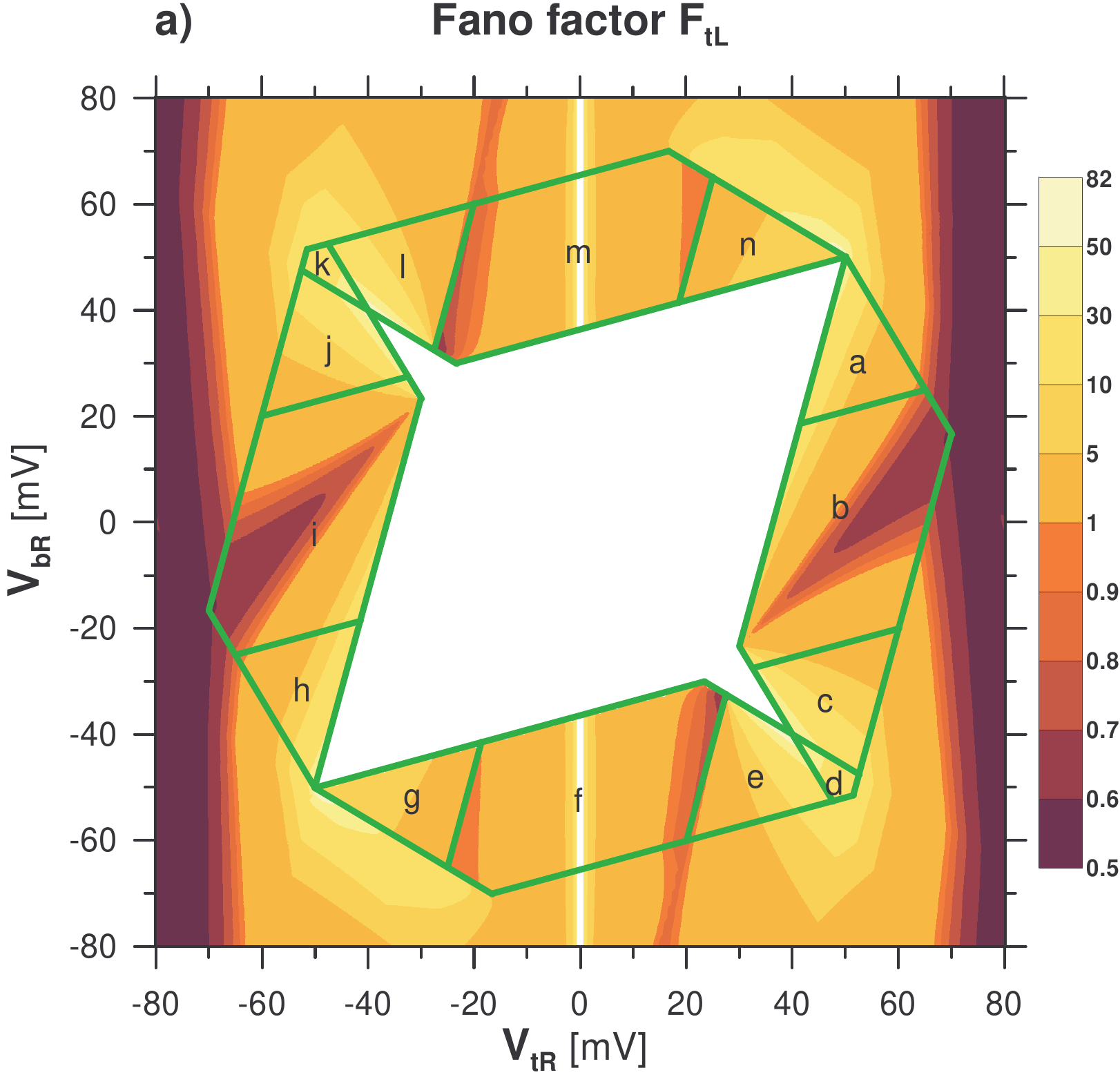}
\includegraphics[width=0.4\textwidth]{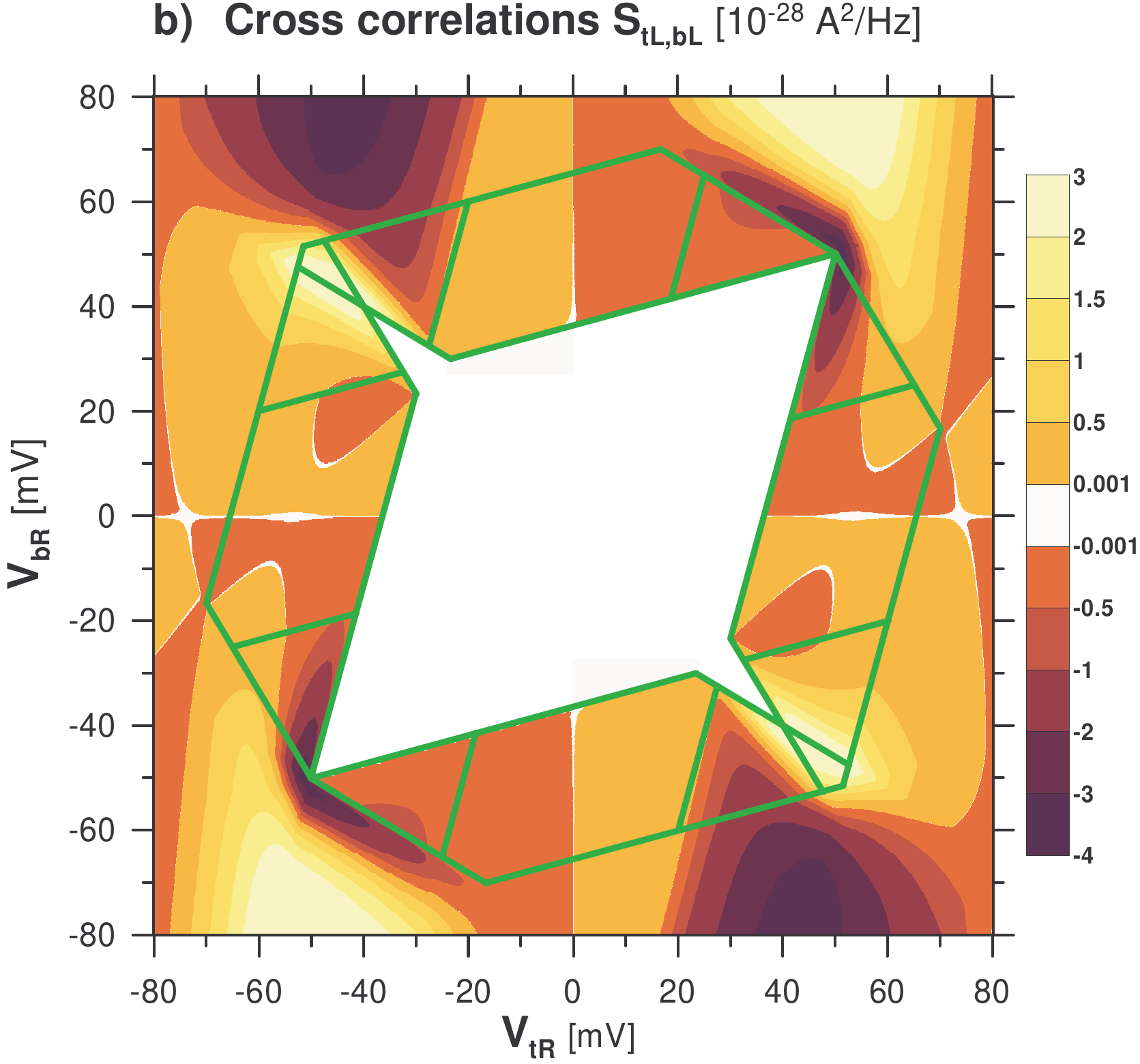}
\caption{(Color online) Maps (a) for the Fano factor $F_{tL}$ and (b) for the cross-correlation $S_{tL,bL}$ in the strong inter-dot case $C_{int}=3$ aF and plotted for $R_{tL}=R_{tR}=1$ M$\Omega$, $R_{bL}=10$ M$\Omega$, $R_{bR}=100$ M$\Omega$. The other parameters are the same as in Fig.~\ref{fig3}.}\label{fig12}
\end{figure}

\begin{figure}
\includegraphics[width=0.4\textwidth]{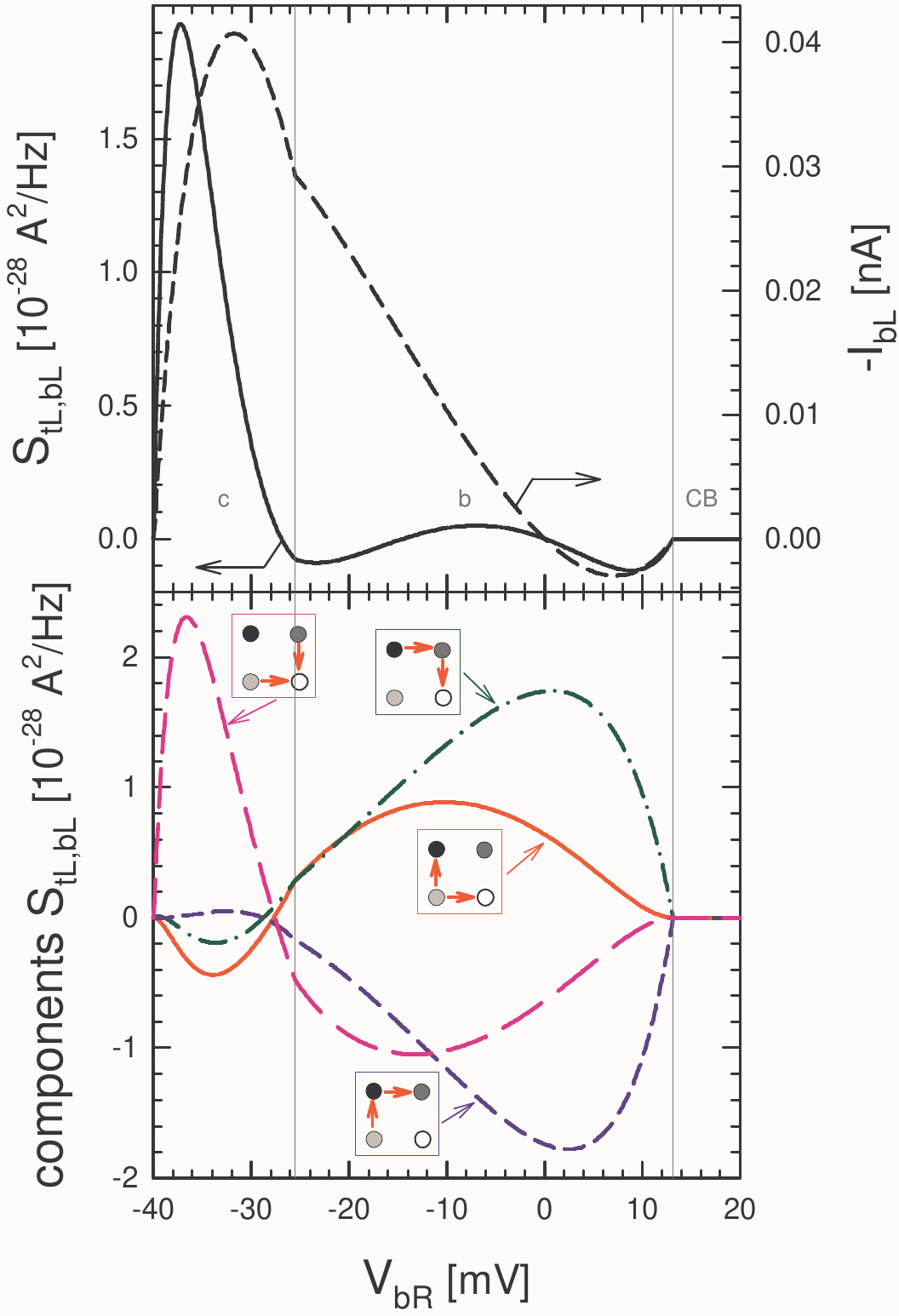}
\caption{(Color online) a) The cross-correlation function $S_{tL,bL}$ (black solid curve) and the current $I_{bL}$ (black short dash curve) in the bottom QD plotted vs. $V_{bR}$ for the cross-section through regions $c$ and $b$ at $V_{tR}=40$ mV. The other parameters are the same as in Fig.~\ref{fig12}. b) The components of $S_{tL,bL}$: the pink long dash curve corresponds to $D_{tL,bL}^{-10,01}$, which describes correlations of the tunneling events $(-1,0)\rightarrow (0,0)$ through the left top tunnel barrier and $(0,1)\rightarrow (0,0)$ through the left bottom tunnel barrier; the red solid curve corresponds to $D_{tL,bL}^{-10,-10}$ and describes correlations between $(-1,0)\rightarrow (0,0)$ and $(-1,0)\rightarrow (-1,1)$; the green dash-dot curve corresponds to $D_{tL,bL}^{-11,01}$ and describes correlations between $(-1,1)\rightarrow (0,1)$ and $(0,1)\rightarrow (0,0)$; and the blue short dash curve corresponds to $D_{tL,bL}^{-11,-10}$, which describes correlations between $(-1,1)\rightarrow (0,1)$ and $(-1,0)\rightarrow (-1,1)$.}\label{fig13}
\end{figure}

A strong correlation can be seen even when one part of the device is unbiased, e.g., for $V_{bR}=0$ and $V_{tR}>V^{th}$ in the region $b$ in Fig.~\ref{fig6}(d). There occurs a pumping effect: charge transfer in the top QD leads to injection (or ejection) of charge in the bottom part of the device. Therefore, one can observe a large enhancement of the shot noise $S_{bL,bL}$, in which the relevant contributions are from the injected $I_{bL}^+$ and the ejected $I_{bL}^-$ current to the Schottky term $S^{Sch}_{bL}$. Our studies resemble shot noise in the presence of Coulomb drag in coupled quantum wires.\cite{Trauzettel-2002,Aguado-2000} In both the systems local charge and voltage fluctuations make electronic transport
strongly correlated,\cite{Trauzettel-2002,Aguado-2000} which is seen in the current-current correlation functions. Of course our system is different, because we have assumed incoherent transport with electron thermalization at each quantum dot and electron transfer between the dots is prohibited, thus a momentum transfer and exchange processes are absent in our system -- in contrast to coupled quantum wires, where these processes are relevant for the Coulomb drag effect.

We showed also that the current cross-correlation function $S_{tL,bL}$ is in general negative, indicating on anti-bunching between charges transferred through the both channels. Let us stress again that here anti-bunching is a dynamical process caused by Coulombic repulsion between transferred electrons (strong potential anti-correlations on the both QDs), but it is not fermionic origin (as in recent studies for electrons scattered in nanostructures\cite{kulik,khlus,lesovik,buttiker1990,blanter}). The function $S_{tL,bL}$ can have a plateau or can be non-monotonic as a function of bias voltage.  The analysis of its components $D_{tL,bL}^{n'_tn'_b,n_tn_b}$ showed that there are negative and positive contributions corresponding to forward and backward tunneling processes. Moreover, it was seen that due to retarded and advanced processes individual components $D_{tL,bL}^{n'_tn'_b,n_tn_b}=0$ at some finite bias voltages. All these various contributions give $S_{tL,bL}$ as a non-monotonic function, which can change its sign in some situations with asymmetric tunneling resistances.

McClure et al.\cite{mcclure07} performed an experiment on a 2QD system, which showed sign reversal of the current cross-correlation function $S_{t,b}$ near a honeycomb vertex in the charge stability space. Applying gate voltages to the top and the bottom QD they could go from one stability region to another. Since the sign of $S_{t,b}$ depends on directions of tunneling processes (whether electrons are injected to or ejected from QDs), sign reversal of $S_{t,b}$ was observed at a state degeneracy line between the stability regions. The effect was also seen when one bias voltage was reversed. Notice that the experiment was performed close to border lines separating different Coulomb blockade regions and bias voltages were small. Our studies are more general and focus on dynamical aspects of current correlations, which occur for large bias voltages. It would be interesting to extend the experimental setup\cite{mcclure07,zhang07} outside the Coulomb blockade regime and verify our theoretical predictions, for example, to see competition of various contributions to the cross-correlation function $S_{tL,bL}$.

We presented also the voltage-voltage correlation functions $S_{V_{\ell},V_{\ell'}}$, which are connected with charge fluctuations in our system. Since nanotechnology and shot noise measurement technique made recently great progress, we believe that such the correlation functions are not purely theoretical interest and they can be measured together with current correlation functions.

\acknowledgments{The work was supported as a part of the European Science Foundation EUROCORES Programme FoNE by funds from the Ministry of Science and Higher Education in years 2006-09).}

\newpage

\section*{References}
\begin{thebibliography}{99}

\bibitem{mcclure07} D. T. McClure, L. DiCarlo, Y. Zhang, H.-A. Engel, C. M. Marcus, M. P. Hanson, and A. C. Gossard, Phys. Rev. Lett. \textbf{98}, 056801 (2007).

\bibitem{zhang07} Y. Zhang, L. DiCarlo, D. T. McClure, M. Yamamoto, S. Tarucha, C. M. Marcus, M. P. Hanson, and A. C. Gossard, Phys. Rev. Lett. \textbf{99}, 036603 (2007).

\bibitem{feynman} R. P. Feynman, R. B. Leighton, and M. Sands, \textit{The Feynman Lectures} (Addison-Wesley, Reading, MA, 1965), Vol. 3.

\bibitem{hbt} R. Hanbury Brown, R. Q. Twiss, Nature \textbf{177}, 27 (1956).

\bibitem{loudon} R. Loudon, Phys. Rev. A \textbf{58}, 4904 (1998).

\bibitem{texier} C. Texier and M. B\"{u}ttiker, Phys. Rev. B \textbf{62}, 7454 (2000).

\bibitem{martin} T. Martin and R. Landauer, Phys. Rev. B \textbf{45}, 1742 (1992).

\bibitem{liu} R. C. Liu, B. Odom, Y. Yamamoto, and S. Tarucha, Nature \textbf{391}, 263 (1998).

\bibitem{oliver} W. D. Oliver, J. Kim, R. C. Liu, and Y. Yamamoto, Science \textbf{284}, 299 (1999).

\bibitem{henny} M. Henny, S. Oberholzer, C. Strunk, T. Heinzel, K. Ensslin, M. Holland, and C. Sch\"{o}nenberger, Science \textbf{284}, 296 (1999).

\bibitem{kulik} I. O. Kulik and A. N. Omel'yanchuk, Fiz. Nizk. Temp. \textbf{10}, 305 (1984) [Sov. J. Low Temp. Phys. \textbf{10}, 158 (1984)].

\bibitem{khlus} V. A. Khlus, Zh. Eksp. Teor. Fiz. \textbf{93}, 2179 (1987) [Sov. Phys. JETP \textbf{66}, 1243 (1987)].

\bibitem{lesovik} G. B. Lesovik, Pis'ma Zh. Eksp. Teor. Fiz. \textbf{49}, 513 (1989) [JETP Lett. \textbf{49}, 592 (1989)].

\bibitem{buttiker1990} M. B\"{u}ttiker, Phys. Rev. Lett. \textbf{65}, 2901 (1990).

\bibitem{blanter} Ya. M. Blanter and M. B\"{u}ttiker, Phys. Rep. \textbf{336}, 1 (2000).

\bibitem{korotkov92} A. N. Korotkov, D. V. Averin, K. K. Likharev, and S. A. Vasenko, in \textit{Single-Electron Tunneling and Mesoscopic Devices}, edited by H. Koch and H. L\"{u}bbig, Springer Series in Electronics and Photonics Vol. 31, (Springer, Berlin, 1992), p. 45.

\bibitem{hershfield} S. Hershfield, J. H. Davies, P. Hyldgaard, C. J. Stanton, and J. W. Wilkins, Phys. Rev. B \textbf{47}, 1967 (1993).

\bibitem{korotkov94} A. N. Korotkov, Phys. Rev. B \textbf{49}, 10381 (1994).

\bibitem{iannaccone} G. Iannaccone, G. Lombardi, M. Macucci, and B. Pellegrini, Phys. Rev. Lett. \textbf{80}, 1054 (1998); Nanotechnology \textbf{10}, 97 (1999).

\bibitem{kuznetsov} V. V. Kuznetsov, E. E. Mendez, J. D. Bruno, and J. T. Pham, Phys. Rev. B \textbf{58}, R10159 (1998).

\bibitem{bb99} B. R. Bu{\l}ka, J. Martinek, G. Micha{\l}ek, and J. Barna\'{s}, Phys. Rev. B \textbf{60}, 12246 (1999).

\bibitem{safonov} S. S. Safonov, A. K. Savchenko, D. A. Bagrets, O. N. Jouravlev, Y. V. Nazarov, E. H. Linfield, and D. A. Ritchie, Phys. Rev. Lett. \textbf{91}, 136801 (2003).

\bibitem{sukhorukov} E. V. Sukhorukov, G. Burkard, and D. Loss, Phys. Rev. B \textbf{63}, 125315 (2001).

\bibitem{thielmann} A. Thielmann, M. H. Hettler, J. K\"{o}nig, and G. Sch\"{o}n, Phys. Rev. Lett. \textbf{95}, 146806 (2005).

\bibitem{wu} S.-T. Wu and S. Yip, Phys. Rev. B \textbf{72}, 153101 (2005).

\bibitem{rychkov} V. Rychkov and M. B\"{u}ttiker, Phys. Rev. Lett. \textbf{96}, 166806 (2006).

\bibitem{oberholzer} S. Oberholzer, E. Bieri, C. Sch\"{o}nenberger, M. Giovannini, and J. Faist, Phys. Rev. Lett. \textbf{96}, 046804 (2006).

\bibitem{burkard} G. Burkard, D. Loss, and E. V. Sukhorukov, Phys. Rev. B \textbf{61}, R16303 (2000).

\bibitem{michalek} G. Micha{\l}ek and B. R. Bu{\l}ka, Eur. Phys. J. B \textbf{28}, 121 (2002).

\bibitem{aghassi08} J. Aghassi, M. H. Hettler, and G. Sch\"{o}n, Appl. Phys. Lett. \textbf{92}, 202101 (2008).

\bibitem{haupt08} S. Haupt, J. Aghassi, M. H. Hettler, and Gerd Sch\"{o}n, arXiv:0802.3579.

\bibitem{hwang07} M.-J. Hwang, M.-S. Choi, and R. Lopez, Phys. Rev. B \textbf{76}, 165312 (2007).

\bibitem{weymann08} I. Weymann, Phys. Rev. B \textbf{78}, 045310 (2008).

\bibitem{bb00} B. R. Bu{\l}ka, Phys. Rev. B \textbf{62}, 1186 (2000).

\bibitem{cottetprb} A. Cottet, W. Belzig, and C. Bruder, Phys. Rev. B \textbf{70}, 115315 (2004).

\bibitem{cottetprl} A. Cottet, W. Belzig, and C. Bruder, Phys. Rev. Lett. \textbf{92}, 206801 (2004).

\bibitem{cottetepl} A. Cottet and W. Belzig, Europhys. Lett. \textbf{66}, 405 (2004).

\bibitem{gustavssonprl} S. Gustavsson, R. Leturcq, B. Simovic, R. Schleser, T. Ihn, P. Studerus, K. Ensslin, D. C. Driscoll, and A. C. Gossard, Phys. Rev. Lett. \textbf{96}, 076605 (2006).

\bibitem{gustavssonprb} S. Gustavsson, R. Leturcq, B. Simovic, R. Schleser, P. Studerus, T. Ihn, K. Ensslin, D. C. Driscoll, and A. C. Gossard, Phys. Rev. B \textbf{74}, 195305 (2006).

\bibitem{schon} G. Sch\"{o}n, in \textit{Quantum Transport and Dissipation}, edited by T. Dittrich, P. H\"{a}nggi, G.-L. Ingold, B. Kramer, G. Sch\"{o}n, and W. Zwerger (Wiley-VCH Verlag, New York, 1998), Chap. 3.

\bibitem{vliet} K. M. van Vliet and J. R. Fassett, in \textit{Fluctuation Phenomena in Solids}, edited by R. E. Burgess (Academic Press, New York, 1965), p. 267.

\bibitem{symmetric} H.-A. Engel and D. Loss, Phys. Rev. Lett. \textbf{93}, 136602 (2004); E. A. Rothstein, O. Entin-Wohlman, and A. Aharony, Phys. Rev. B \textbf{79}, 075307 (2009).

\bibitem{bb08} B. R. Bu{\l}ka, Phys. Rev. B \textbf{77}, 165401 (2008).

\bibitem{norm} D. S\'{a}nchez and R. L\'{o}pez, Phys. Rev. B \textbf{71}, 035315 (2005); S. V. Vaseghi, \textit{Advanced Digital Signal Processing and Noise Reduction} (Willey, New York, 2000), Chap. 3.4.7.

\bibitem{Trauzettel-2002} B. Trauzettel, R. Egger, and H. Grabert, Phys. Rev. Lett. \textbf{ 88}, 116401 (2002).

\bibitem{Aguado-2000} R. Aguado and L. P. Kouwenhoven, Phys. Rev. Lett. \textbf{84}, 1986 (2000); E. Onac, F. Balestro, L. H. Willems van Beveren, U. Hartmann, Y. V. Nazarov, and L. P. Kouwenhoven, Phys. Rev. Lett. \textbf{96}, 176601 (2006).

\end{thebibliography}
\end{document}